\documentclass{aastex631}

\usepackage{booktabs}

\newcommand{\sw}[1]{\texttt{#1}}

\begin{document}

\title{AT2023vto: An Exceptionally Luminous Helium Tidal Disruption Event from a Massive Star}

\author[0000-0003-0871-4641]{Harsh Kumar}
\affiliation{Center for Astrophysics \textbar{} Harvard \& Smithsonian, 60 Garden Street, Cambridge, MA 02138-1516, USA}
\affiliation{The NSF AI Institute for Artificial Intelligence and Fundamental Interactions, USA}

\author[0000-0002-9392-9681]{Edo Berger}
\affiliation{Center for Astrophysics \textbar{} Harvard \& Smithsonian, 60 Garden Street, Cambridge, MA 02138-1516, USA}
\affiliation{The NSF AI Institute for Artificial Intelligence and Fundamental Interactions, USA}

\author[0000-0002-1125-9187]{Daichi Hiramatsu}
\affiliation{Center for Astrophysics \textbar{} Harvard \& Smithsonian, 60 Garden Street, Cambridge, MA 02138-1516, USA}
\affiliation{The NSF AI Institute for Artificial Intelligence and Fundamental Interactions, USA}

\author[0000-0001-6395-6702]{Sebastian Gomez}
\affiliation{Space Telescope Science Institute, 3700 San Martin Drive, Baltimore, MD 21218, USA}

\author[0000-0003-0526-2248]{Peter K.~Blanchard}
\affiliation{Center for Astrophysics \textbar{} Harvard \& Smithsonian, 60 Garden Street, Cambridge, MA 02138-1516, USA}

\author[ 0000-0001-7007-6295]{Yvette Cendes}
\affiliation{Center for Astrophysics \textbar{} Harvard \& Smithsonian, 60 Garden Street, Cambridge, MA 02138-1516, USA}
\affiliation{Department of Physics, University of Oregon, Eugene, OR 97403, USA}

\author{K. Azalee Bostroem}  
\affiliation{Las Cumbres Observatory, 6740 Cortona Drive, Suite 102, Goleta, CA 93117-5575, USA} 
\author{Joseph Farah}   
\affiliation{Las Cumbres Observatory, 6740 Cortona Drive, Suite 102, Goleta, CA 93117-5575, USA} 
\affiliation{Department of Physics, University of California, Santa Barbara, CA 93106-9530, USA}
\author{Estefania Padilla Gonzalez}  
\affiliation{Las Cumbres Observatory, 6740 Cortona Drive, Suite 102, Goleta, CA 93117-5575, USA}
\affiliation{Department of Physics, University of California, Santa Barbara, CA 93106-9530, USA}
\author{D. Andrew Howell}   
\affiliation{Las Cumbres Observatory, 6740 Cortona Drive, Suite 102, Goleta, CA 93117-5575, USA} 
\affiliation{Department of Physics, University of California, Santa Barbara, CA 93106-9530, USA}
\author{Curtis McCully} 
\affiliation{Las Cumbres Observatory, 6740 Cortona Drive, Suite 102, Goleta, CA 93117-5575, USA} 
\author{Megan Newsome}
\affiliation{Las Cumbres Observatory, 6740 Cortona Drive, Suite 102, Goleta, CA 93117-5575, USA} 
\affiliation{Department of Physics, University of California, Santa Barbara, CA 93106-9530, USA}
\author{Giacomo Terreran} 
\affiliation{Las Cumbres Observatory, 6740 Cortona Drive, Suite 102, Goleta, CA 93117-5575, USA}

\begin{abstract}
We present optical/UV observations and the spectroscopic classification of the transient AT2023vto as a tidal disruption event (TDE) at $z=0.4846$. The spectrum is dominated by a broad \ion{He}{2} $\lambda 4686$ emission line, with a width of $\approx 3.76\times 10^4$ km s$^{-1}$ and a blueshift of $\approx 1.05\times 10^4$ km s$^{-1}$, classifying it as a member of the TDE-He class.  The light curve exhibits a long rise and decline timescale, with a large peak absolute magnitude of $M_g\approx -23.6$, making it the most luminous of the classical optical TDEs (H, H+He, He) discovered to date by about 2 mag (and $\approx 4$ mag compared to the mean of the population). The light curve exhibits a persistent blue color of $g-r\approx -0.4$ mag throughout its evolution, similar to other TDEs, but distinct from supernovae. We identify the host galaxy of AT2023vto in archival Pan-STARRS images and find that the transient is located at the galaxy center, and that its inferred central black hole mass is $\sim 10^7$ M$_\odot$. Modeling the light curves of AT2023vto, we find that it resulted from the disruption of a $\approx 9$ M$_\odot$ star by a $\approx 10^7 M_\odot$ supermassive black hole. The star mass is about 5 times larger than the highest star masses previously inferred in TDEs, and the black hole mass is at the high end of the distribution. AT2023vto is comparable in luminosity and timescale to some putative TDEs (with a blue featureless continuum), as well as to the mean of the recently identified population of ambiguous nuclear transients (ANTs), although the latter are spectroscopically distinct and tend to have longer timescales. ANTs have been speculated to arise from tidal disruptions of massive stars, perhaps in active galactic nuclei, and AT2023vto may represent a similar case but in a dormant black hole, thereby bridging the TDE and ANT populations. We anticipate that Rubin Observatory / LSST will uncover similar luminous TDEs to $z\sim 3$.

\end{abstract}


\section{Introduction} 
\label{sec:intro}

Over the past decade, the observed population of tidal disruption events (TDEs; \citealt{1988Natur.333..523R, 1989ApJ...346L..13E}) has grown rapidly to tens of events. The observed TDE population appears to be dominated by the disruption of stars spanning $\sim 0.1-2$ M$_\odot$ by supermassive black holes (SMBHs) spanning $\sim 10^{5.5}-10^{7.5}$ M$_\odot$ \citep{2019ApJ...872..151M, 2021ApJ...908....4V, 2022MNRAS.515.5604N, 2023ApJ...942....9H, 2023ApJ...949..113G, 2023ApJ...955L...6Y}. Spectroscopic observations have also demonstrated a diversity of TDEs, with spectra dominated by broad hydrogen lines (TDE-H), a mix of hydrogen and helium (TDE-H+He) and only helium lines (TDE-He); e.g. \citet{2014ApJ...793...38A, 2021ARA&A..59...21G, 2022MNRAS.515.5604N, 2022A&A...659A..34C, 2023ApJ...942....9H, 2023ApJ...955L...6Y}. These emission lines often exhibit velocity shifts, indicating outflowing gas driven by intense radiation pressure near the black hole, with velocities ranging from hundreds to $\sim 10^4$ km s$^{-1}$ \citep{2019ApJ...879..119H, 2019MNRAS.488.1878N, 2022A&A...659A..34C, 2022A&A...666A...6W}.  Thus TDEs provide an excellent laboratory for the study of accretion and outflows around otherwise dormant SMBHs, and a census of their demographics.

Along with the growing population of spectroscopically-classified (``classical'') optical TDEs, other putative or potentially related events have also been recognized in wide-field optical surveys.  These include events with similar luminosities, timescales, colors, and an origin in host galaxy nuclei, but lacking any spectral features (termed ``featureless'' TDEs; \citealt{2023ApJ...942....9H}), as well as events with a contentious origin such as ASASSN-15lh, which has been argued to be either a TDE \citep{2016NatAs...1E...2L} or a superluminous supernova \citep{2016Sci...351..257D, 2016ApJ...828....3B}, or the very luminous transient AT2021lwx, whose origin is unclear but has been speculated to potentially be a TDE \citep{2023ApJ...948L..19S}. Relatedly, a new class of nuclear transients, dubbed ambiguous nuclear transients \citep[ANTs;][]{2011ApJ...735..106D, 2017NatAs...1..865K, 2020MNRAS.494.2538N, 2022ApJ...933..196H, 2024arXiv240611552W}, has also been recognized.  ANTs are loosely defined as transients coinciding with their host galaxy nuclei, which do not resemble spectroscopically supernovae, TDEs, or normal active galactic nuclei.  Some ANTs are $\sim 1-2$ orders of magnitude more luminous than classical TDEs, and exhibit much longer durations \citep{2024arXiv240508855H,2024arXiv240611552W}, and it has been speculated that they may represent tidal disruptions of massive stars, much beyond the typical solar mass stars in classical TDEs \citep{2024arXiv240611552W}; in this scenario it remains unclear why disruptions of such massive stars would lead to distinct optical spectra.

Against this backdrop, here we present the study of the most luminous classical TDE to date -- AT2023vto -- with a peak luminosity exceeding the median of the TDE population by nearly two orders of magnitude, but comparable to the median for the ANT population, and which exhibits a broad and blueshifted \ion{He}{2} line characteristic of the TDE-He class, and distinct from ANTs.  The high luminosity of AT2023vto points to the disruption of a massive star ($\sim 10$ $M_\odot$), and it potentially represents a bridge between the classical TDE and luminous ANT populations.  The paper is arranged as follows.  In \S\ref{sec:obs}, we summarize the discovery and multi-wavelength follow-up observations of AT2023vto and its host galaxy. In \S\ref{sec:TDESLSN} we demonstrate that AT2023vto is a a classical TDE. In \S\ref{sec:model}, we model the UV/optical light curve to determine the properties of AT2023vto, and show that it resides in a unique part of the star mass -- black hole mass phase-space of TDEs.  We compare the observed and inferred properties of AT2023vto to the TDE population and other potentially related phenomena (e.g, ANTs) in \S\ref{sec:disc}, and summarize the key results and implications in \S\ref{sec:conclusion}.

\section{Discovery and Observations} 
\label{sec:obs}

\subsection{Discovery}

AT2023vto was first detected on 2023 September 9 by ZTF (internal name: ZTF23abcvbqq) at R.A.=$00^{h} 24^{m} 34.71^{s}$, Decl.=$+47^{\circ} 13^{\prime} 21.55^{\prime\prime}$ (J2000) with a magnitude of $m_g = 20.98\pm 0.33$~\citep{2023TNSTR2714....1F}. On 2023 November 21, \citet{2023TNSCR3050....1P} obtained an optical spectrum of the transient and classified it as a Type II superluminous supernova (SLSN-II) at $z=0.48463$, based on its luminosity and the tentative identification of a faint H$\beta$ emission feature. Below, we show that while the redshift is correct, the identification of H$\beta$ is likely erroneous (or the spectrum has evolved significantly), and AT2023vto is not a SLSN-II, but a TDE.

\subsection{Host Galaxy}
\label{sec:host}

\begin{figure}[t!]
\begin{center}
    \includegraphics[width=0.8\linewidth]{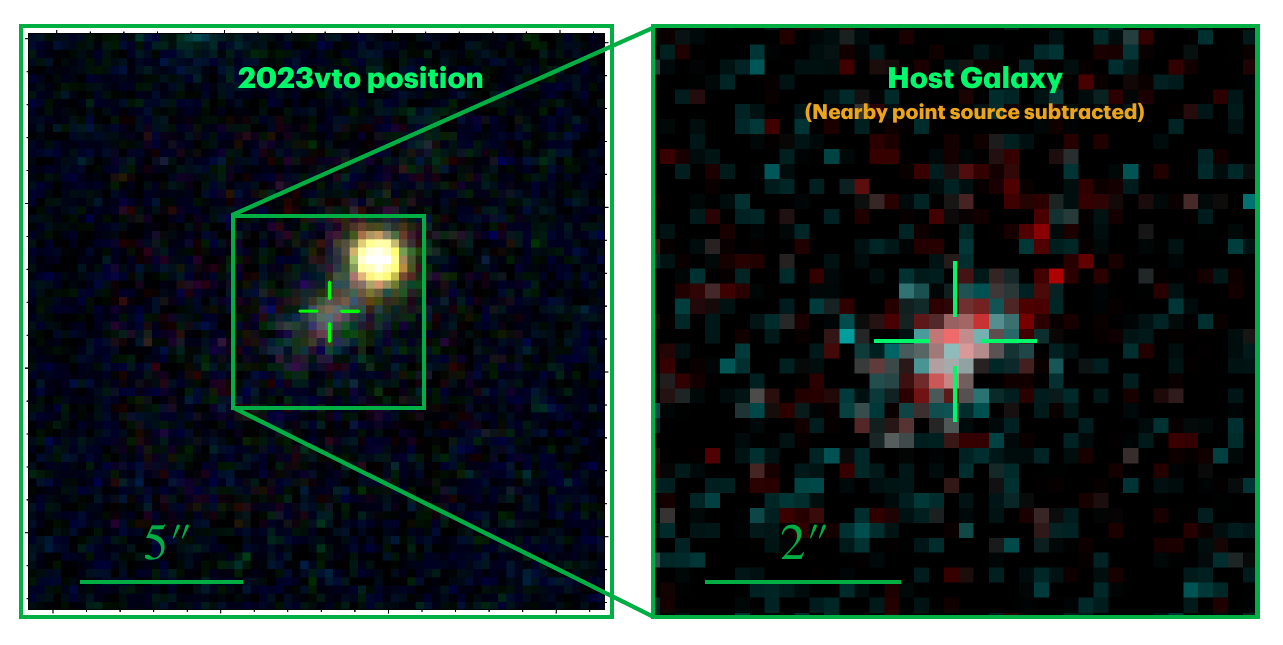}
\end{center}
\caption{Archival Pan-STARRS-DR1 images centered on the location of AT2023vto. There are two distinct sources within a few arcseconds of the location of the transient; the fainter extended source directly underlies the position of AT2023vto and is its host galaxy (the brighter point source is a likely foreground star). {\it Right:} A zoomed-in image with the point source subtracted. We find that AT2023vto coincides with the center of its host galaxy, supporting its TDE origin.
\label{fig:gal}}
\end{figure}

Inspection of the Pan-STARRS Data Release 1 (PS-DR1) archival images reveals two sources within a few arcseconds of the location of AT2023vto; see Figure~\ref{fig:gal}. The brighter, point-like source is well resolved from the location of AT2023vto (offset by $2.3^{\prime \prime}$), while the fainter, extended source directly underlies the position of AT2023vto.  As shown in Figure~\ref{fig:gal}, subtracting the nearby source using the PSF profile determined from the images results in clean residuals, implying it is a likely foreground star. We identify the extended source as the host galaxy of AT2023vto, with the following magnitudes determined from the PS-DR1 images after subtracting the nearby point source: $m_{g} = 22.60 \pm 0.18$, $m_{r} = 21.33 \pm 0.16$, $m_{i} = 20.54 \pm 0.22$, $m_{z} = 19.99\pm 0.23$, and $m_{y} = 20.03 \pm 0.19$. This indicates that contamination from host galaxy light in UV and blue optical bands is negligible. At the redshift of the galaxy, the corresponding absolute magnitudes are $M_{g} = -20.07 \pm 0.18$, $M_{r} = -21.34 \pm 0.16$, $M_{i} = -22.13 \pm 0.22$, $M_{z} =  -22.68 \pm 0.23$, and $M_{y} = -22.64 \pm 0.19$.

\begin{figure}[t!]
\begin{center}
    \includegraphics[width=0.9\linewidth]{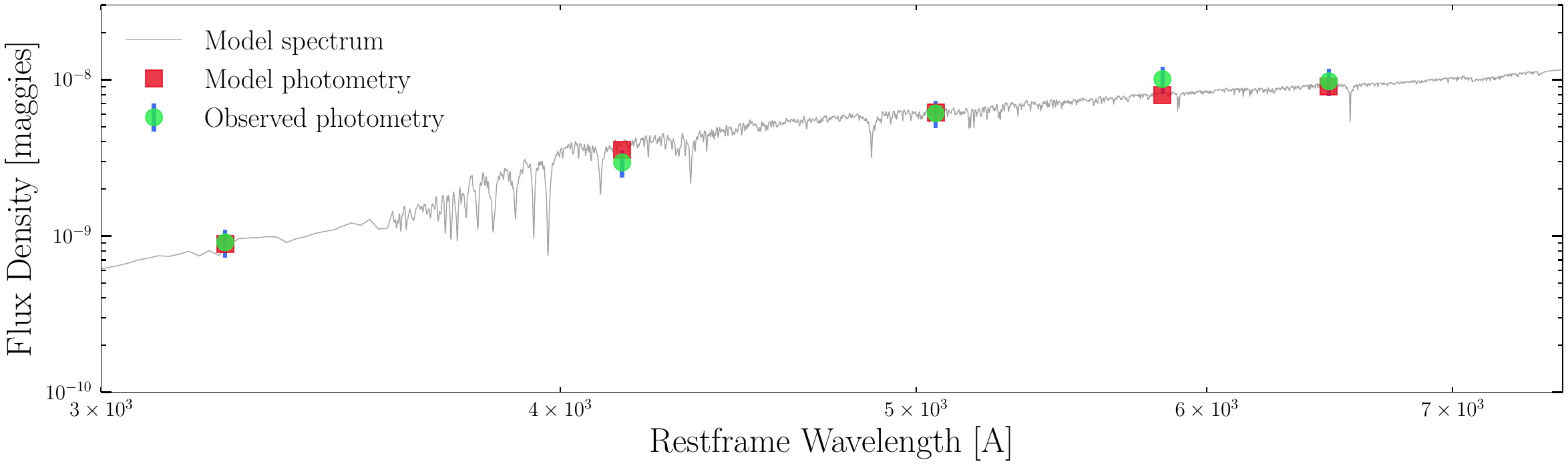}
\end{center}
\caption{Spectral energy distribution of the host galaxy of At2023vto (green points) along with the best-fit model from \sw{Prospector} (red points and grey curve). The host photometry has been measured from PanSTARRS-DR1 images with the nearby point source subtracted (see Figure~\ref{fig:gal}).
\label{fig:galprosfit}}
\end{figure}

We model the host spectral energy distribution with the \sw{Prospector} code \citep{2017ApJ...837..170L} to estimate its stellar mass and hence to infer its associated SMBH mass.  We use an exponential star formation history and include the metallicity and dust contributions as free parameters. The resulting best-fit SED model is shown in Figure~\ref{fig:galprosfit}. We find a stellar mass of ${\rm log}(M_{\rm gal}/M_\odot) = 10.74\pm 0.24$ and a metallicity of ${\rm log}(Z) = -0.11^{+0.38}_{-0.28}$. Using the galaxy stellar mass to black hole mass relation of ${\rm log}(M_{\rm BH}/M_{\odot}) = 6.70 + 1.61 \times {\rm log}(M_{\rm gal}/ 3 \times 10^{10} M_{\odot})$ \citep{2020ARA&A..58..257G, 2021ApJ...907...77Z}, we estimate a black hole mass of ${\rm log}(M_{\rm BH}/M_\odot)= 7.12\pm 0.38$.

\subsection{Imaging Observations}

Motivated by the initial SLSN-II classification, we began follow-up observations of AT2023vto with the Sinistro cameras on the network of Las Cumbres Observatory (LCO; \citealt{2013PASP..125.1031B}) 1-m telescopes in the $U,B,V,g,r$ bands, through the Global Supernova Project \citep{2017AAS...23031803H}. LCO photometry was performed with point-spread function (PSF) fitting using \texttt{lcogtsnpipe}\footnote{\url{https://github.com/LCOGT/lcogtsnpipe}} \citep{2016MNRAS.459.3939V}, a PyRAF-based photometric reduction pipeline. The $U,B,V$-band data were calibrated to Vega \citep{2000PASP..112..925S} magnitudes using standard fields observed on the same night by the same telescope as AT2023vto, while the $g,r,i$-band data were calibrated to AB magnitudes using the 9th Data Release of the AAVSO Photometric All-Sky Survey \citep{2016yCat.2336....0H}. All photometric measurements in the LCO $g,r$ bands have been obtained after image subtraction with the in-house zogy algorithm python pipeline \citep{2016ApJ...830...27Z}.  We did not perform image subtraction in the $U,B,V$ bands due to negligible host contamination in these blue bands.

\begin{figure}[t!]
\centering\includegraphics[width=0.95\linewidth]{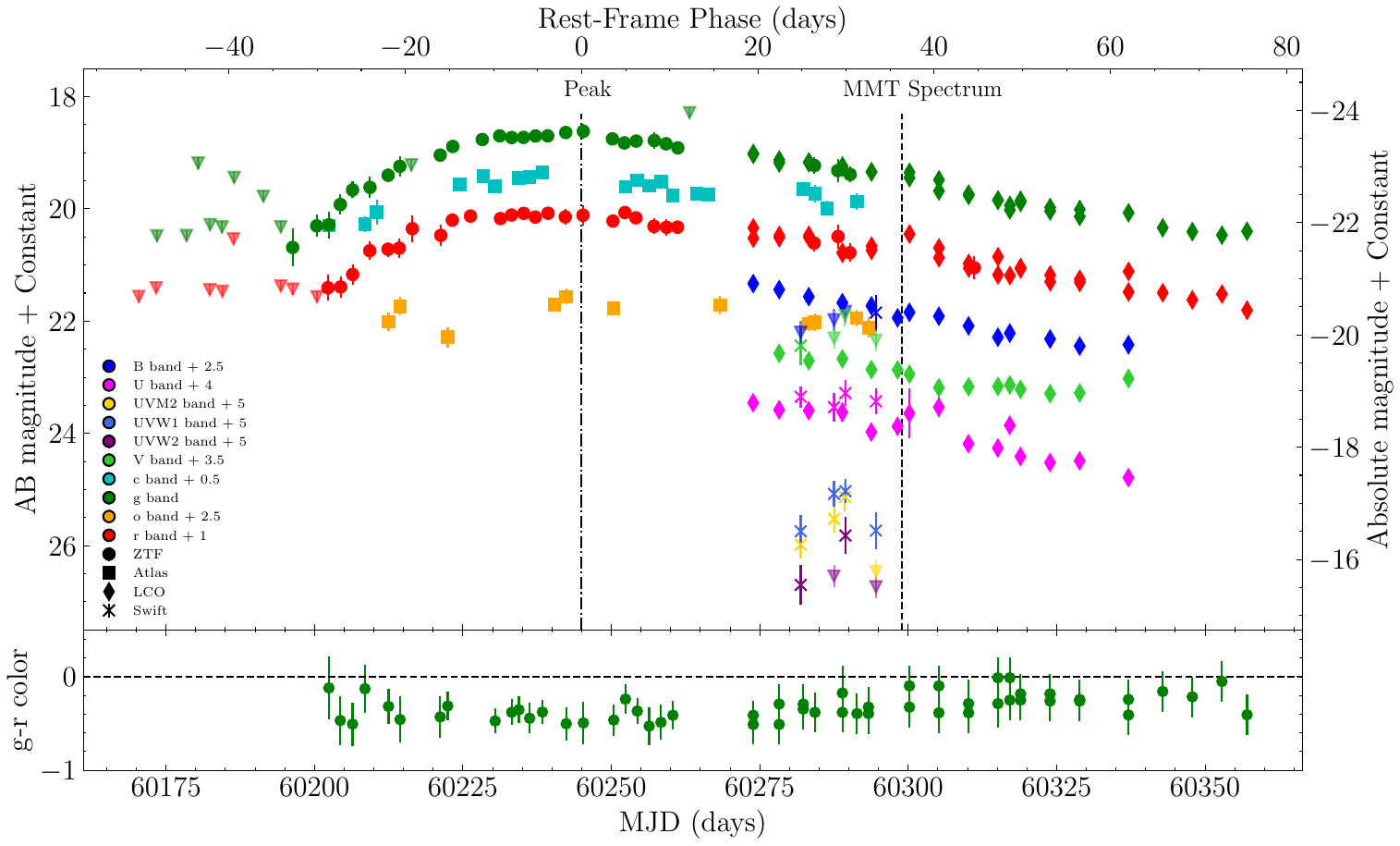}
\caption{Optical/UV light curves of AT2023vto. The magnitudes are corrected for Galactic Extinction. Photometry has been obtained from ZTF (circles), ATLAS (squares), LCO (diamonds), and {\it Swift}  (crosses). AT2023vto peaks in $g$-band about 50 days post-discovery with a peak absolute magnitude of $m_g\approx -23.6$. The $g-r$ color is blue ($\approx -0.4$ mag) and does not evolve significantly during the rise and decline phase of the light curve for about 155 days, typical of TDEs.
\label{fig:lc}}
\end{figure}

We additionally obtained publicly available data from ZTF \citep{2014htu..conf...27B, 2019PASP..131a8002B}, ATLAS~\citep{2018PASP..130f4505T}, and the Neil Gehrels {\it Swift} Observatory \citep{2004ApJ...611.1005G}. We obtained the ZTF $g,r$-band data through the Automatic Learning for the Rapid Classification of Events broker~\citep[\sw{ALeRCE;}][]{2021AJ....161..242F}. The data span from 2023 September 9 to December 12 in $g$-band, and from 2023 September 15 to 2024 January 2 in the $r$-band.

ATLAS observed AT2023vto in the $c,o$-bands, and we obtained the photometry from the ATLAS forced Photometry Server~\citep{2021TNSAN...7....1S}, selecting all detections with a signal-to-noise ratio of $\geq5$. To reduce scatter from the individual exposures, we determined a weighted median of all observations in 2-day bins. The data span 2023 September 15 to December 13 in $c$-band, and 2023 September 25 to December 15, in $o$-band.
 
{\it Swift} observations were obtained with the UV/Optical Telescope (UVOT; \citealt{2005SSRv..120...95R}) in all six filters (UVW2, UVM2, UVW1, $U$, $B$, and $V$). The observation commenced on 2023 December 3 (Target ID 16397), with four visits up to December 16. We obtained three additional observations from 2024 May 25 to June 2, for reference templates and performed aperture photometry with \texttt{Swift\_host\_subtraction}\footnote{\url{https://github.com/gterreran/Swift_host_subtraction/tree/main}}, following the standard procedures described in \cite{2009AJ....137.4517B,2014Ap&SS.354...89B}.

All photometric measurements reported in this work are in AB magnitudes and corrected for Galactic extinction, with $E(B - V) = 0.088$ mag \citep{2011ApJ...737..103S}, assuming the \citet{1999PASP..111...63F} reddening law with $R_V = 3.1$. Given the very blue colors and lack of host galaxy \ion{Na}{1} D absorption lines in the spectrum (see below), the host galaxy extinction is expected to be negligible; this is verified independently with \sw{MOSFiT} modeling in \S~\ref{sec:model}.

The resulting multi-band light curve is shown in Figure~\ref{fig:lc}. We find that the light curve exhibits a slow rise for $\approx 50$ d post-discovery, rising by $\approx 2$ mag in $g$-band and peaking at $m_g = 18.61 \pm 0.13$, or M$_g\approx -23.6$, mag on MJD = 60245.3 (which we define as phase = 0). The light curve then declines slowly, taking $\approx 50$ d to reach half of its peak brightness. Equally important, the light curve exhibits a persistent blue $g-r$ color of $-0.43\pm 0.10$ mag throughout its evolution (with a possible slight reddening beyond MJD of about 60,300).

\subsection{Optical Spectroscopy}

We obtained an optical spectrum of AT2023vto with the Binospec spectrograph \citep{2019PASP..131g5004F} on the MMT 6.5-m telescope on 2023 December 21, corresponding to an observed phase of $\approx 54$ d (rest-frame phase of $\approx 37$ d). We analyzed the spectrum using standard IRAF routines in the {\tt twodspec} package. The spectrum was bias-subtracted and flat-fielded, the sky background was modeled and subtracted from the 2D image, and the one-dimensional spectra were optimally extracted, weighing by the inverse variance of the data. A wavelength calibration was applied using an arc lamp spectrum taken directly after the science image. Relative flux calibration was applied to the spectrum using a standard star taken close to the observation time.

\begin{figure}[t!]
\centering\includegraphics[width=0.95\linewidth]{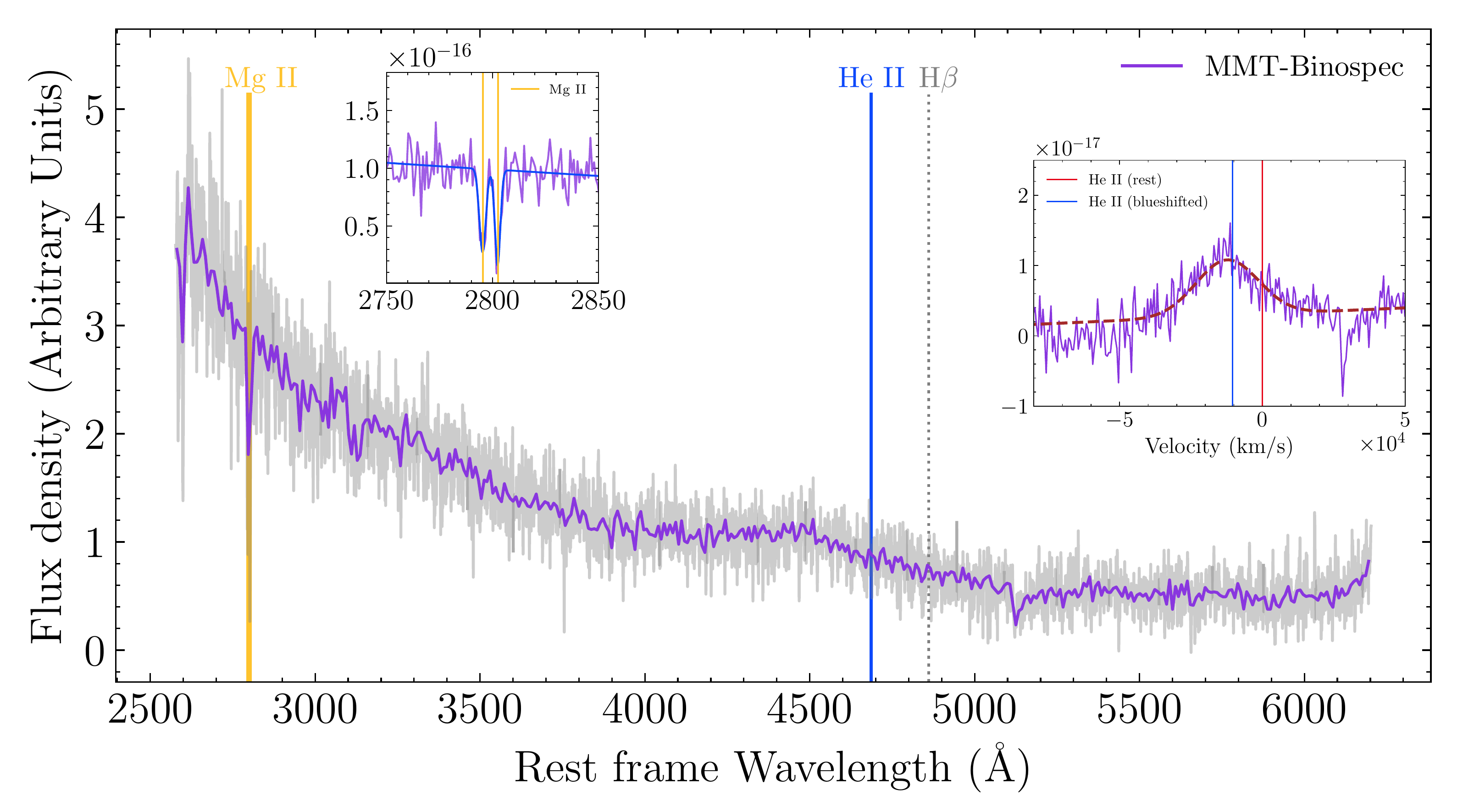}
\caption{MMT/Binospec optical spectrum of AT2023vto obtained at a phase of +54 days, shown in rest-frame wavelength (the purple line is binned by a factor of 5). The narrow \ion{Mg}{2} absorption doublet indicates a redshift of $z=0.48463\pm 0.00004$ (left inset). We find a single broad emission feature centered at $4524$ \AA, with $v_{\rm FWHM}\approx 3.76\times 10^4$ km s$^{-1}$ and a blueshift of $\approx 1.05\times 10^4$ km s$^{-1}$ relative to the rest wavelength of \ion{He}{2}$\lambda 4686$ (right inset).  This feature is inconsistent with H$\beta$ (dotted vertical line), indicating a TDE-He classification.
\label{fig:specevo}}
\end{figure}

The spectrum, shown in Figure~\ref{fig:specevo}, reveals a strong \ion{Mg}{2} absorption doublet from the host galaxy, which gives a redshift of $z=0.48463\pm 0.00004$ (see inset of Figure~\ref{fig:specevo}). Additionally, we detect a broad emission feature centered at a rest-frame wavelength of $\approx 4524$ \AA, which corresponds most closely to \ion{He}{2} $\lambda 4686$. Since the earlier spectrum used for the initial classification  (observed phase of 26 days; \citealt{2023TNSCR3050....1P}) is not publicly available, we cannot determine if the broad feature we observe was misidentified as H$\beta$, or whether the spectrum has evolved from being dominated by H$\beta$ to \ion{He}{2} $\lambda 4686$; such a transition was claimed for the TDE AT2017eqx \citep{2019MNRAS.488.1878N}, but over a much longer timescale of $\approx 100$ d compared to only $\approx 30$ d in this case.  Regardless, based on our MMT spectrum, we re-classify AT2023vto as a TDE-He event.

\begin{figure}[t!]
\center\includegraphics[width=0.9\linewidth]{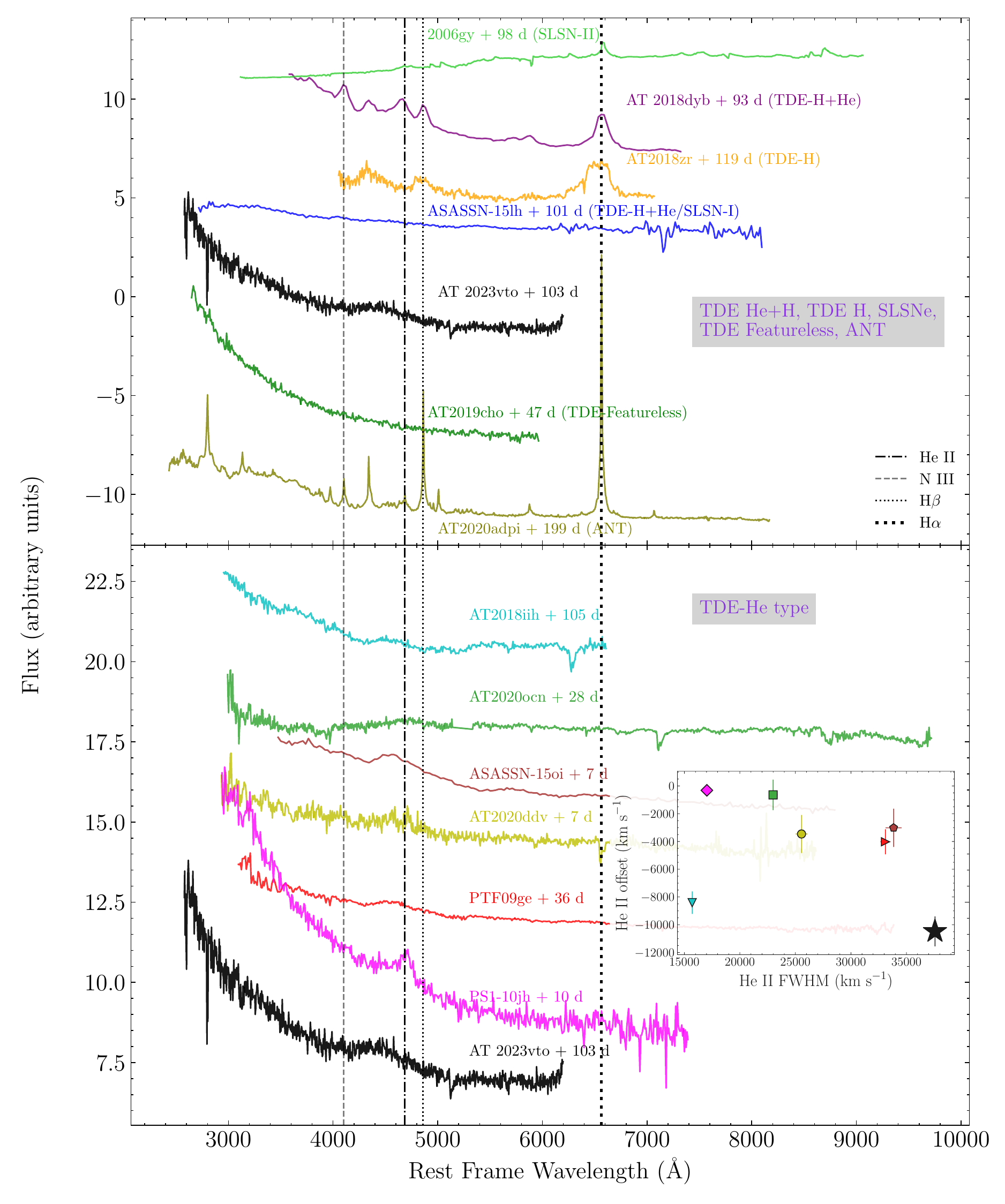}
\caption{{\it Top:} The optical spectrum of AT2023vto compared to all TDE types (H, H+He, He, ``featureless''), SLSN-II, ANT, and ASASSAN-15lh; the quoted phases are relative to the discovery date of each event. {\it Bottom:} A comparison to all other TDE-He events found to date. AT2023vto exhibits a bluer continuum compared to the other TDE-He (matched only by PS1-10jh, but at a much earlier phase of its evolution).  The width and blueshift of the \ion{He}{2} $\lambda 4686$ line of AT2023vto compared to the other TDE-He events is shown in the inset (colors match those of the spectra in the main panel).  AT2023vto exhibits the largest velocity width and blueshift, although neither quantity is completely unprecedented in previous TDE-HE events.
\label{fig:speccomp}}
\end{figure}

Fitting the emission feature with a Gaussian profile, we find that it peaks at $\lambda = 4524\pm 16$ \AA, with a full-width at half maximum of $v_{\rm FWHM}=(3.76\pm 0.01)\times 10^4$ km s$^{-1}$ (see inset in Figure~\ref{fig:specevo}). The velocity shift of the line center relative to the rest wavelength is $v_{\rm He} = (-1.05\pm 0.11)\times 10^4$ km s$^{-1}$, which is large but not unprecedented for TDE-He (\citealt{2019ApJ...879..119H}; see \S\ref{sec:disc}).  If the observed feature was H$\beta$, the resulting velocity shift would be $\approx -2.1\times 10^4$ km s$^{-1}$, which is an order of magnitude larger\footnote{Or in Type II SNe~\citep{2014MNRAS.441..671A}.} than observed in other TDEs \citep{2022A&A...659A..34C}.  A blackbody fit to the continuum (with the emission feature subtracted) gives a temperature of $T_{\rm BB}=19,926\pm 211$~K.  We note that this is likely a lower limit as the spectral peak is blueward of the observed wavelength range of our spectrum.

\subsection{Radio Observations}

Following our identification of AT2023vto as a TDE, we obtained radio observations with the Karl G.~Jansky Very Large Array (VLA) on 2024 June 19 ($284$ days post optical discovery) using Director's Discretionary Time (Program ID 24A-479, PI: Cendes). We observed AT2023vto in C-band ($4-8$ GHz), using the primary calibrator 3C147, and the secondary calibrator J2355+4950.  We processed the VLA data using standard data reduction procedures in the Common Astronomy Software Application package (CASA; \citealt{McMullin2007}), using {\tt tclean} on the calibrated measurement set available in the NRAO archive, with Briggs weighting. We do not detect a radio source at the location of AT2023vto, with a $3\sigma$ upper limit of $F_\nu\approx 16.5$ $\mu$Jy, corresponding to a luminosity limit of $\nu L_\nu\lesssim 9\times10^{38}$ erg s$^{-1}$. This limit is not constraining in the context of TDEs with non-relativistic outflows \citep{Alexander2020,Cendes2023}, but rules out the luminosities seen in TDEs with on-axis relativistic jets \citep{p1,p2,Brown2017}.  Future observations will determine if AT2023vto exhibits late-time radio emission as seen in a substantial fraction of TDEs \citep{Cendes2023}.

\subsection{X-ray Observations}

We analyzed \textit{Swift} X-Ray Telescope (XRT; $0.3-10$\,keV) observations obtained along with the UVOT observations (phase of $32-49$ d). No emission was detected at the location of AT2023vto, and we estimate a $3\sigma$ upper limit using the \textit{Swift}-XRT web tool\footnote{\url{https://www.swift.ac.uk/user_objects/index.php}} \citep{2007A&A...469..379E,2009MNRAS.397.1177E}. With a Milky Way H~{\sc i} column density of $7.63\times10^{20}$\,cm$^{-2}$ \citep{2016A&A...594A.116H}\footnote{Via the NASA HEASARC N$_{\rm H}$ Tool: \url{https://heasarc.gsfc.nasa.gov/cgi-bin/Tools/w3nh/w3nh.pl}} and assuming a power-law spectrum with a photon index of 2, the stacked source count rate limit is converted\footnote{Via the NASA HEASARC WebPIMMS: \url{https://heasarc.gsfc.nasa.gov/cgi-bin/Tools/w3pimms/w3pimms.pl}} to an unabsorbed flux limit of $F_{\rm X}\lesssim 9.2 \times 10^{-13}$ erg s$^{-1}$ cm$^{-2}$, corresponding to a luminosity limit of $L_{\rm X}\lesssim 8.7 \times 10^{44}$ erg s$^{-1}$. This limit is not particularly constraining relative to the majority of TDEs, which have soft X-ray luminosities of $\sim 10^{41} - 10^{44}$ erg s$^{-1}$ \citep{2017ApJ...838..149A, 2021ARA&A..59...21G}.

\section{AT2023vto is a Tidal Disruption Event} \label{sec:TDESLSN}

AT2023vto was classified initially as a SLSN-II based on the claimed detection of H$\beta$ emission \citep{2023TNSCR3050....1P}.  We have shown that our optical spectrum is instead dominated by \ion{He}{2} $\lambda 4686$, and no H$\beta$ emission is detected. The velocity width of the emission feature, $\approx 3.8\times 10^4$ km s$^{-1}$, is much broader than typically observed in SNe, but is more in line with velocities observed in TDE spectra \citep{2022A&A...659A..34C, 2022MNRAS.510.5426P}; see Figure~\ref{fig:speccomp}. The line center is blueshifted relative to the rest-frame of \ion{He}{2} by $\approx 1.05\times 10^4$ km s$^{-1}$, which is  higher than in previous TDE-He events, but at least one other event has a comparable blueshift (Figure~\ref{fig:specevo}). 

Another strong indication of a TDE origin is the persistent and blue $g-r\approx -0.4$ mag color of AT2023vto during both the rise and decline phases of the light curve. Such persistent blue colors are observed in other TDEs \citep{2023ApJ...942....9H}, with a typical range of about $-0.5$ to $0$ mag (mean of $\approx -0.4$ mag).  On the other hand, this is in contradiction to Type IIn SNe or SLSNe-II, which show a typical rapid transition from blue color on the rise, to a red color ($g-r\gtrsim 0$ mag) post-peak (e.g., \citealt{2018MNRAS.475.1046I,2024ApJ...964..181H}).

Finally, we consider the location of AT2023vto relative to its underlying host galaxy. The centroid position of the galaxy lies at the same pixel (pixel size = $0.25^{\prime \prime}$) in PS1 as the centroid location of AT2023vto. We performed relative astrometry between the LCO and PanSTARRS DR1 images to determine the offset between AT2023vto and its host galaxy. We find a negligible offset of $0.13\pm 0.12^{\prime \prime}$; see Figure~\ref{fig:gal}. The offset uncertainty of $0.12^{\prime \prime}$ corresponds to a projected physical distance uncertainty of $\approx 0.7$ kpc at the redshift of AT2023vto.

To summarize, the spectroscopic identification of a broad \ion{He}{2} emission line, combined with the long duration, large peak luminosity, persistent blue $g-r$ color, and a location consistent with the nucleus of its host galaxy, all indicate that AT2023vto is a TDE-He event.

\section{Light Curve Modeling} 
\label{sec:model} 

Having identified AT2023vto as a TDE, we now turn to modeling its light curves with the \sw{MOSFiT} Python package, which uses a Markov chain Monte Carlo (MCMC) method to model the light curves of transients powered by various energy sources \citep{2017ascl.soft10006G}. The \sw{MOSFiT} TDE model calculates the luminosity of the system by converting the fallback accretion rate of the disrupted stellar material to radiation (via an efficiency parameter). The model includes several parameters: the mass of the black hole (M$_{\mathrm{BH}}$); the mass of the disrupted star (M$_{*}$); the scaled impact parameter ($b$), accounting for the star's polytropic index and the bound mass fraction such that a value of $b = 1$ implies a full disruption; the efficiency of fallback accretion to radiation conversion ($\epsilon$); the photosphere power-law constant ($R_{\rm ph,0}$) and exponent ($l$); and the viscous time ($T_{viscous}$), which governs the formation rate of the accretion disk. Additionally, the model assumes a blackbody spectral energy distribution to determine the brightness in each band. The model is based on \citet{2019ApJ...872..151M} where further details are provided.  To model the observed light curves, we used the \sw{emcee} ensemble sampler~\citep{2013PASP..125..306F} with 200 walkers and 20,000 steps to ensure convergence of the model with a potential scale reduction factor (PSRF) of $<1.1$.

\begin{figure}[t!]
\center\includegraphics[width=0.95\linewidth]{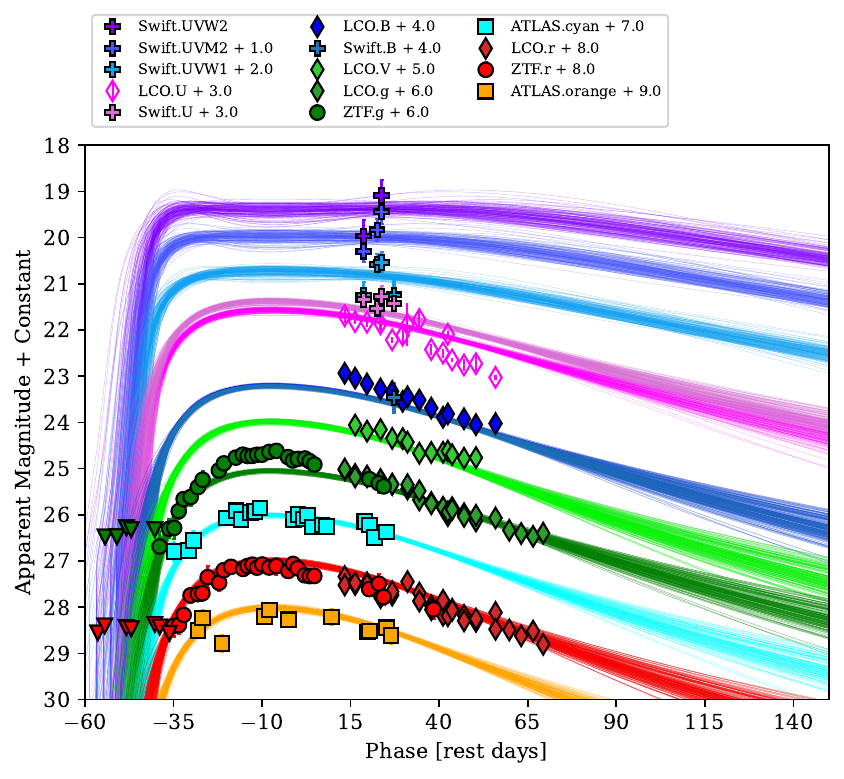}
\caption{Multi-band \sw{MOSFiT} TDE model light curves for AT2023vto. We note that the LCO $U$-band data (open symbols) were not used in the fit, but their inclusion does not change the resulting model parameters. The best-fit model parameters are listed in Table~\ref{table:parameters}.
\label{fig:modelfit}}
\end{figure}

\begin{table}[h!]
\caption{Posterior Distribution of the TDE Model Parameters from \sw{MOSFiT}.}
\centering
\begin{tabular}{lccc}
\toprule
Parameter & Range & log-prior & Posterior \\
  % &   &   &  (without {\it Swift} data)  \\ %& (with {\it Swift} data) \\
\midrule

$M_*$ (M$_\odot$) & [0.01, 100] & False  & $9.07^{+2.98}_{-2.65}$   \\

$M_{\mathrm{BH}}$ (M$_\odot)$ &  [$10^4$, $10^{10}$] & True & $6.98^{+0.06}_{-0.05}$ \\

$t_{\rm exp}$ (days) &  [$-100$, 0] & False & $-19.7^{+4.1}_{-7.2}$ \\

$b$ &  [0,2] & False  & $0.53^{+0.08}_{-0.06}$   \\

$\epsilon$ & [$5\times 10^{-3}$, 0.4]  & True & $-2.23^{+0.12}_{-0.06}$ \\

$ T_{\rm viscous}$ (days) & [0.01, $10^3$] & True &  $-1.27^{+1.3}_{-1.19}$  \\

$ R_{\rm ph,0}$  & [$10^{-4}$, $10^4$] & True & $0.97^{+0.06}_{-0.05}$ \\

$l_{\rm ph}$ &  [0,4] & False & $1.22^{+0.20}_{-0.13}$ \\

$n_{H,\rm host}$ (cm$^{-2}$) &  [$10^{16}$, $10^{23}$]  & True  & $17.98^{+1.32}_{-1.13}$ \\

$ A_{\rm V}$ (mag) &   & False  & $0.00^{+0.01}_{-0.00}$\\

$\sigma$ & [$10^{-3}$, 100]  & True & $-0.75^{+0.03}_{-0.03}$ \\

\bottomrule
\end{tabular}
\label{table:parameters}
\end{table} 

In Figure~\ref{fig:modelfit}, we plot the best-fit model light curve realizations in the optical and UV bands, and we list the resulting best-fit parameters with $1\sigma$ confidence intervals in Table~\ref{table:parameters}. The model indicates that AT2023vto resulted from the disruption of an $\approx 9.1$ M$_\odot$ star by a $\approx 10^7$ M$_\odot$ black hole.  The latter is in good agreement with the estimated value from the host galaxy stellar mass (\S\ref{sec:host}). The estimated disruption time is $\approx 20$ d prior to the initial ZTF detection. The normalized impact parameter is $b\approx 0.53$, and the efficiency is $\epsilon\approx 0.006$.  As expected, the best-fit host galaxy extinction is consistent with zero.

The model provides an overall good fit to the optical and UV data.  We note that it does not fully capture the light curve behavior near the peak in $g$-band (the model is overall smoother than the data).  This is potentially due to the simplified nature of the TDE model; for example, the models are calibrated for lower mass stars than inferred for AT2023vto, and it is possible that high mass stars have somewhat different internal structures, which will affect the debris dynamics during disruption \citep{2015ApJ...798...64G, 2013sse..book.....K}. We similarly note that the model overpredicts the observed brightness in the LCO $U$-band. This is likely due to the impact of the atmospheric cutoff in the ground-based $U$-band data, which is not included in the \sw{MOSFiT} model and is difficult to correct for given the unusually blue color of AT2023vto; we did not include this filter in the model fitting\footnote{We found that inclusion of the $U$-band data has a negligible effect on the inferred parameters, for example $\lesssim 1\%$ in the values of $M_*$ and $M_{\rm BH}$.}.

The best fit model from \sw{MOSFiT} results in a peak bolometric luminosity of $L_{p,\rm bol} = (7.7\pm 0.4) \times 10^{44}$ erg s$^{-1}$, a blackbody temperature of $T_{p,\rm BB} = 13,560 \pm 378$ K and a blackbody radius of $R_{p,\rm BB} = (5.6 \pm 0.3) \times 10^{15}$ cm. The radiated energy integrated from the estimated disruption time to the time of the last available observation is $E_{\rm rad}\approx 1.8 \times 10^{52}$ erg.

\section{Discussion}
\label{sec:disc}

\subsection{Comparison to the Light Curves and Models of Optical TDEs}
\label{sec:comp}

In this section, we compare AT2023vto to the optical TDE population, and to other possibly related transients, using its observed light curve and spectroscopic properties, as well as the inferred physical parameters from \sw{MOSFiT} modeling compared to those of the optical TDE population (Gomez S., at el., in prep.)

\begin{figure}[t!]
\begin{center}
    \includegraphics[width=0.475\linewidth]{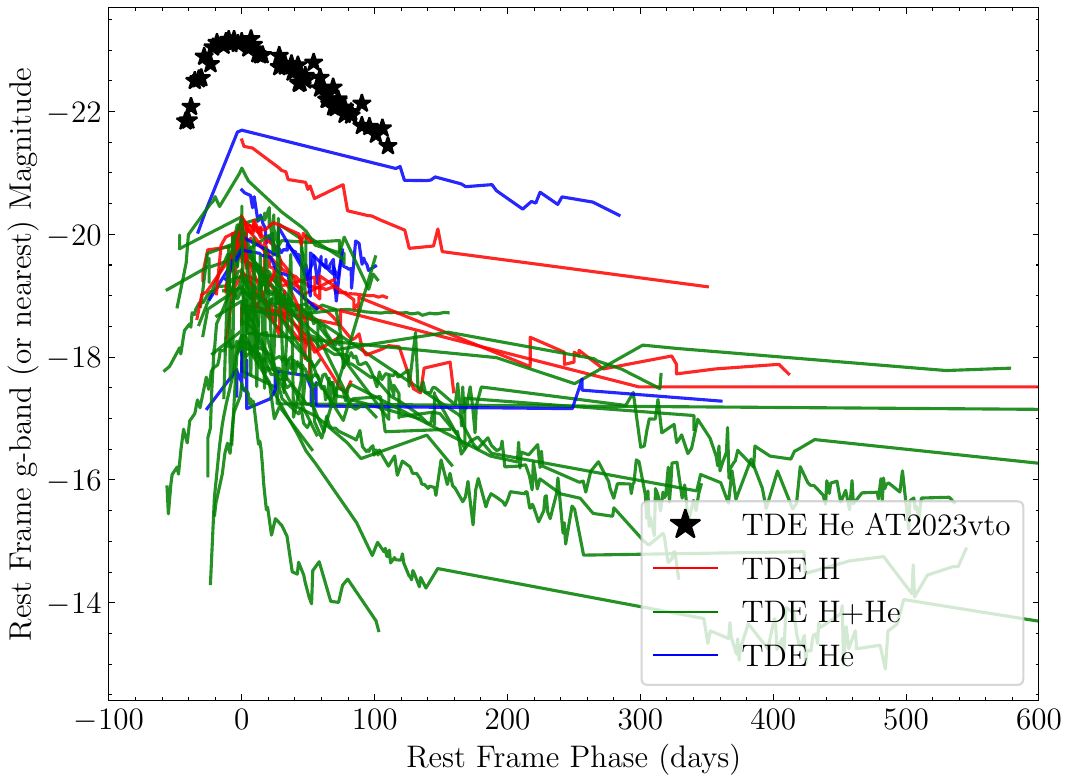}\hfill
    \includegraphics[width=0.475\linewidth]{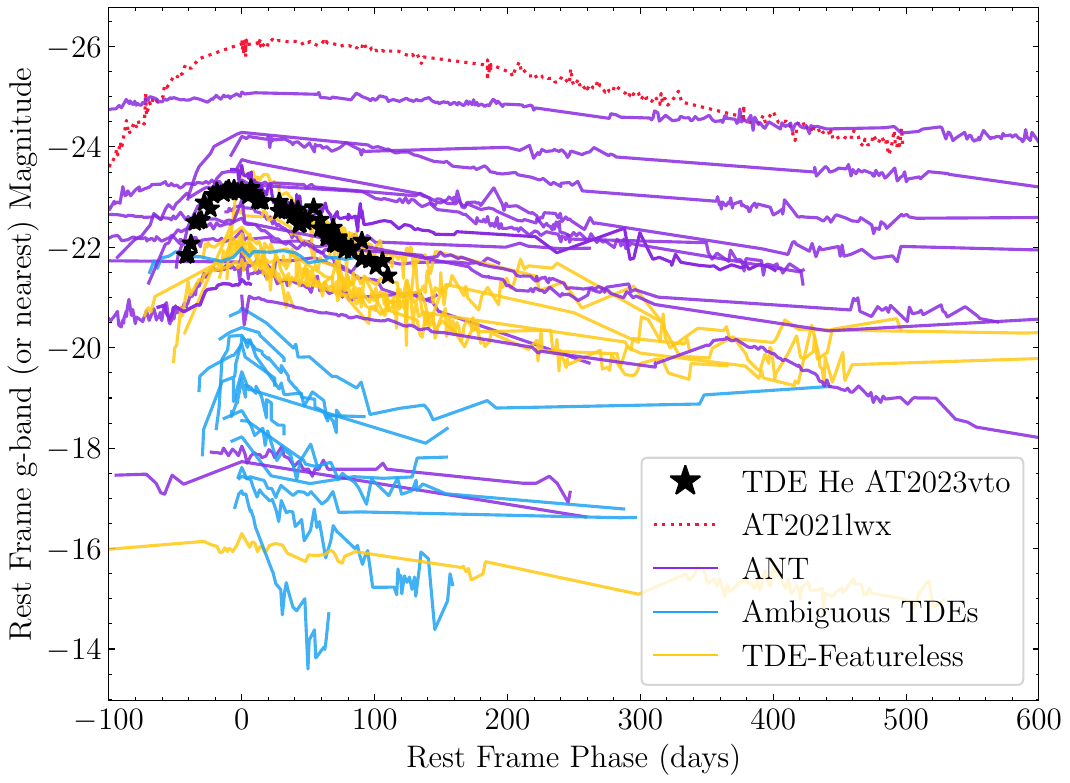}
\end{center}
\caption{{\it Left:} A comparison of the rest-frame $g$-band light curve of AT2023vto (black stars) to the classical TDE sample, consisting of TDE-H (red), TDE-H+He (green) and TDE-He (blue). AT2023vto is the brightest among this population of spectroscopically-classified TDEs, outshining the second brightest event by $\approx 1.5$ mag, and the median by $\approx 4$ mag. {\it Right:} A comparison of AT2023vto's rest-frame $g$-band light curve to putative or possible TDEs, including featureless events, ambiguous events, and ANTs. AT2023vto is at the bright end of the featureless TDEs (yellow), and at the mean of the ANT population. Comparison data from~\citet{2006ApJ...653L..25G, 2014ApJ...780...44C, 2014ApJ...792...53V, 2014ApJ...793...38A, 2014MNRAS.445.3263H, 2016MNRAS.455.2918H, 2016MNRAS.463.3813H, 2016Sci...351..257D, 2017ApJ...842...29H, 2017ApJ...843..106B, 2017ApJ...844...46B, 2017MNRAS.465L.114W, 2018ApJ...852...72V, 2019ApJ...883..111H, 2019MNRAS.488.1878N, 2020MNRAS.497.1925G, 2020MNRAS.498.4119S, 2020MNRAS.499..482N, 2020TNSCR3228....1A, 2020TNSTR2500....1P, 2020TNSTR3332....1N, 2021ApJ...908....4V, 2021ApJ...917....9H, 2021MNRAS.504..792C, 2021TNSCR1632....1Y, 2021TNSCR3611....1Y,
2021TNSTR3208....1F, 2022A&A...666A...6W, 2022ApJ...925...67L, 2022ApJ...937....8Y, 2022MNRAS.515.1146R, 2022MNRAS.515.5604N, 2023ApJ...942....9H, 2023ApJ...949..113G, 2023ApJ...955L...6Y, 2023MNRAS.522.5084G, 2023TNSCR2004....1Y, 2023arXiv231003782S, 2024arXiv240611552W, 2024arXiv240508855H, 2024MNRAS.531.2603H}
\label{fig:comp}}
\end{figure}

In Figure~\ref{fig:comp} we compare the rest-frame $g$-band light curve of AT2023vto to a sample of 72 optical TDEs and potential TDEs (see references in the caption of Figure~\ref{fig:comp}); for the comparison objects, we use the nearest available filter to rest-frame $g$-band. We divide the comparison sample into two categories: (i) classical TDEs, where the TDE classification is based on spectroscopic features (e.g., H and/or He lines); and (ii) putative or potential TDEs, where the spectra are either featureless, or exhibit spectral features that are distinct from classical TDEs (e.g., narrow emission lines as in ANTs), but which are consistent with a nuclear origin. The latter category includes events with featureless spectra that are classified as TDE due to their persistent blue colors, TDE-like light curve properties, and coincidence with the nuclei of their host galaxies~\citep{2021SSRv..217...63V, 2022MNRAS.515.5604N, 2023ApJ...942....9H}; the ambiguous event ASASSAN-15lh, which was claimed to be both a SLSN and a TDE \citep{2016Sci...351..257D, 2016NatAs...1E...2L, 2016ApJ...828....3B}; ANTs, which spatially coincides with their host galaxy nuclei but are spectrally distinct from TDEs \citep{2017NatAs...1..865K,2024arXiv240611552W, 2024arXiv240508855H}; and ambiguous events, which include a variety of transients with a nuclear origin that are referred to as ambiguous/unknown in the TDE literature~\citep{2021SSRv..217...63V, 2022MNRAS.515.5604N, 2023ApJ...942....9H, 2023ApJ...955L...6Y}.

We find that AT2023vto is the brightest among all of the classical spectroscopically-classified TDEs, outshining the second brightest event by $\approx 1.5$ mag, and the median of the population ($-19.7\pm 0.9$ mag) by $\approx 4$ mag. On the other hand, AT2023vto exhibits a similar rise time ($\approx 50$ d) to the optical TDE sample (with $\approx 30-60$ d), as well as a similar decline rate (Figure~\ref{fig:comp}).

Next, we compare the light curve of AT2023vto to the putative TDEs and ANTs. AT2023vto is more luminous than all of the ``ambiguous'' TDE events, and also exceeds the luminosities of the ``featureless'' events, although some of those are comparable in brightness.  On the other hand, AT2023vto is comparable to the {\it mean} of the ANT population, but those generally have longer durations and extend about 2 mag brighter. 

\begin{figure}[t!]
\begin{center}
    \includegraphics[width=0.75\linewidth]{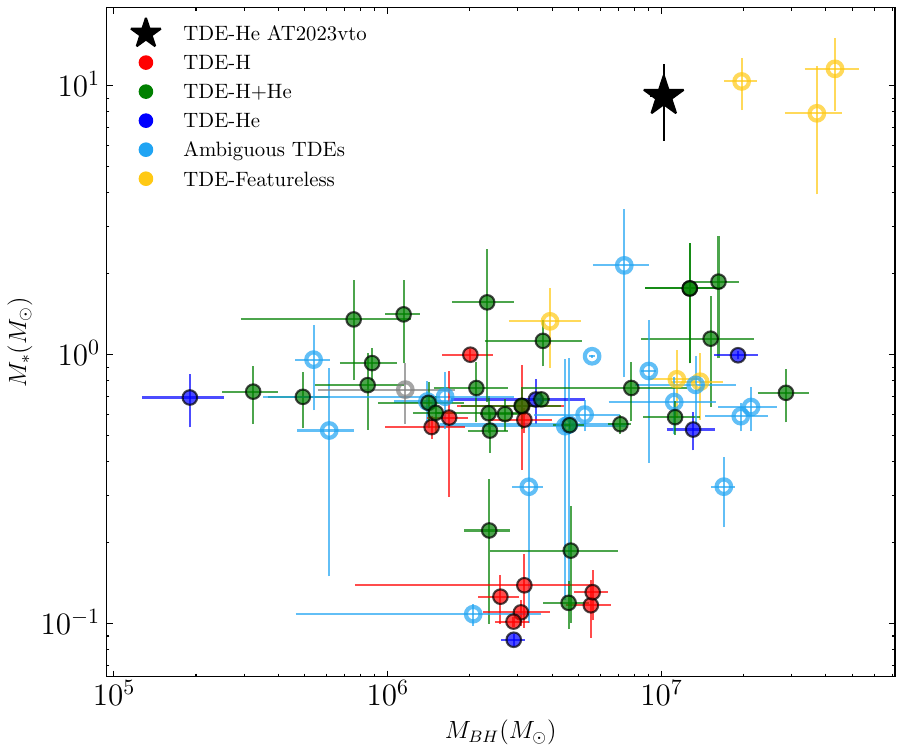}
\end{center}
\caption{Inferred disrupted star mass ($M_{\mathrm{*}}$) versus black hole mass ($M_{\mathrm{BH}}$) from \sw{MOSFiT} model fits of AT2023vto (black star) and optical TDEs, including the H (red), H+He (green), and He (purple) categories, as well as featureless (open yellow) and ambiguous events (open blue).  AT2023vto stands out in this parameter space with an unusually high stellar mass and relatively large black hole mass. A few of the featureless TDEs seem to reside in the same region of parameter space as AT2023vto, and indeed exhibit similar light curves (right panel of Figure~\ref{fig:comp}).
\label{fig:paramdist}}
\end{figure}

In Figure~\ref{fig:paramdist}, we compare the key inferred physical parameters of AT2023vto --- $M_*$ and $M_{\rm BH}$ --- to the classical TDE sample and featureless possible TDEs.  The comparison events are modeled in exactly the same way as our \sw{MOSFiT} model for AT2023vto. As expected from its high luminosity, we find that AT2023vto stands out in terms of the disrupted star mass of $M_{*}\approx 9.1$ M$_\odot$ from the rest of the TDE population, which spans $M_*\approx 0.1-2$ M$_\odot$ with a mean of $\langle M_{*}\rangle = 0.64 \pm 0.46$ M$_\odot$. We also find that $M_{\rm BH}\approx 10^7$ M$_\odot$ in AT2023vto is at the massive end of the distribution, which spans $M_{\rm BH}\approx 10^{5.5}-10^{7.5}$ M$_\odot$, with a mean of $\langle M_{\rm BH}\rangle\approx 10^{6.5}$ M$_\odot$.  For the black hole mass in AT2023vto, its peak bolometric luminosity corresponds to $\approx 0.65$ L$_{\rm Edd}$, which is in the upper range for optical TDEs \citep{2017ApJ...842...29H, 2017ApJ...843..106B, 2019ApJ...872..151M, Wong_2022}. Finally, the scaled impact parameter in AT2023vto, $b\approx 0.55$, is somewhat lower than for other TDE-He events, with $b\sim 1$ \citep{2022MNRAS.515.5604N}.

While AT2023vto resides in a distinct part of the $M_*-M_{\rm BH}$ phase-space than classical TDEs, some of the featureless events (AT2018jbv, AT2020acka, AT2020riz) occupy a similar parameter space, with $M_*\sim 10$ M$_\odot$ and $M_{\rm BH}\sim 10^{7.5}$ M$_\odot$.  Similarly, some of the similarly, or even more luminous, ANTs than AT2023vto, if interpreted as TDEs, have been argued to result from massive stars, $M_*\sim 4.5-90$ M$_\odot$, disrupted by similarly massive black holes, $M_{\rm BH}\sim 10^{7.2}-10^{8.7}$ M$_\odot$ \citep{2024arXiv240611552W}.

\subsection{Comparison to the Spectra of Optical TDEs} 
\label{sec:speccomp}

In Figure ~\ref{fig:speccomp}, we compare the optical spectrum of AT2023vto to those of other TDE-He events, as well as to those of ambiguous or potential TDEs (e.g., ASASSN-15lh, ANTs); we compare these at the nearest epoch to $\sim 100$ d post-discovery available on \sw{WISeREP} \citep{2021TNSAN.142....1Y} to match the epoch of the AT2023vto spectrum.

The spectrum of AT2023vto is broadly similar to those of the other TDE-He events, with the closest analogue being PS1-10jh (albeit at an earlier epoch of about $10$ d post-discovery). AT2023vto has a bluer continuum than the other TDE-He.  In the inset of Figure ~\ref{fig:speccomp}, we plot the \ion{He}{2} $\lambda 4686$ line width versus blueshift for AT2023vto and the 6 comparison TDE-He events.  We find that AT2023vto exhibits the largest line width and the largest blueshift, although some TDE-He events have comparable line widths ($\approx 3\times 10^4$ km s$^{-1}$) or blueshift ($\approx 8\times 10^3$ km s$^{-1}$). We note that none of the TDE-He events exhibit a redshift of the \ion{He}{2} line, as opposed to the trend seen in the Balmer lines of TDE-H and TDE-H+He events \citep{2022A&A...659A..34C}.

While AT2023vto is comparable in peak luminosity to the mean of the ANT population, the latter have clearly distinct spectra compared to AT2023vto and classical TDEs, marked by much narrower Balmer lines, as well as narrow lines of \ion{O}{1}, [\ion{O}{2}] and [\ion{O}{3}] \citep{2024arXiv240508855H, 2024arXiv240611552W}. The Balmer line widths are $\lesssim 3000$ km s$^{-1}$ compared to $\gtrsim 10,000$ km s$^{-1}$ for classical TDEs. \citet{2024arXiv240611552W} found that eight out of ten ANTs in their sample show a narrow [\ion{O}{3}] $\lambda 5007$ line, which is commonly present in AGNs and star-forming galaxy spectra and is indicative of low-density ionized gas. One possibility is that ANTs occur in pre-existing AGNs, explaining their narrower lines from the AGN environment. On the other hand, the spectral properties of AT2023vto match those of the classical TDE-He population, despite a potential similarity in disrupted star mass and black hole mass (\S\ref{sec:comp}), suggesting that it arose from a quiescent SMBH.

\section{Conclusions
\label{sec:conclusion}}

We presented photometric and spectroscopic observations of AT2023vto, a luminous transient at $z = 0.4846$ which we have shown to be a TDE based on the identification of a broad \ion{He}{2}$\lambda 4686$ emission line, persistent $g-r\approx -0.4$ mag color during the long rise and decline of the light curve, and a location consistent with the nucleus of its underlying host galaxy. A comparison to the spectra of other TDEs indicates that AT2023vto is a TDE-He event. We reach the following conclusions based on the observed properties of AT2023vto, and the modeling of its light curves:

\begin{itemize}

\item The peak absolute magnitude of $m_g\approx -23.6$ makes AT2023vto the most luminous classical TDE observed to date, by $\approx 1.9$ mag compared to the second brightest TDE, and by $\approx 3.9$ mag ($\approx 4$ standard deviations) compared to the median of the TDE population.

\item The peak brightness of AT2023vto is similar to some putative TDEs, namely some ``featureless'' events and the mean of the ANT population. However, its TDE-He spectrum is distinct from those of ANTs (which are dominated by narrow lines, including Balmer lines) and the featureless events (which lack any spectroscopic features).

\item The \ion{He}{2} $\lambda 4686$ line in AT2023vto exhibits the largest combination of  velocity width ($\approx 3.76\times 10^4$ km s$^{-1}$) and blueshift ($-1.05\times 10^4$ km s$^{-1}$) of any TDE-He event to date, but neither value on its own is unprecedented.

\item Modeling of the light curves indicates that AT2023vto was caused by the disruption of a $\approx 9.1$ M$_\odot$ star by a $\approx 10^7$ M$_\odot$ black hole (the black hole mass is in good agreement with the value inferred based on the host galaxy mass). This combination is unique amongst optical TDEs; the star mass is about 5 times larger than the most massive stars in previous TDEs, and about $14$ times larger than the mean of the population ($\langle M_{*}\rangle\approx 0.64$). Similar parameters are only found in models of the putative ``featureless'' TDEs, as well as by TDE model fits for ANTs \citep{2022ApJ...933..196H, 2024arXiv240508855H,2024arXiv240611552W, 2024ApJ...964..181H}.

\end{itemize}

AT2023vto is a classical TDE in terms of its observed photometric and spectroscopic properties while having been caused by the disruption of a much more massive star than the general TDE population, similar to those inferred in the TDE model fits to luminous ANTs. Thus, AT2023vto may be a link between the two populations.  The majority of ANTs exhibit different spectra than classical TDEs, dominated by narrow Balmer and oxygen lines.  It is possible that both AT2023vto and the luminous ANTs are caused by massive star disruptions, with the latter occurring in active galactic nuclei and AT2023vto occurring in a dormant SMBH.

The large luminosity and inferred star mass in AT2023vto provide an opportunity to study the extreme end of the TDE luminosity function and the impact of massive star disruptions. With the upcoming Rubin Observatory / LSST, we anticipate that AT2023vto-like events will be detectable to $z\sim 3$ and would, therefore, also probe quiescent SMBHs to much larger redshifts than the bulk of the TDE population. Along with an increase in the detection rate and characterization of ANTs, we expect that such observations will shed light on the connection between luminous TDEs and ANTs.

\begin{acknowledgments}

We thank Brenna Mockler and Conor Ransome for useful discussions. The Berger Time-Domain research group at Harvard is supported by the NSF and NASA grants. The LCO supernova group is supported by NSF grants AST-1911151 and AST-1911225.

Observations reported here were obtained at the MMT Observatory, a joint facility of the Smithsonian Institution and the University of Arizona. This paper uses data products produced by the OIR Telescope Data Center, supported by the Smithsonian Astrophysical Observatory.

This work makes use of observations from the Las Cumbres Observatory global telescope network. The authors wish to recognize and acknowledge the very significant cultural role and reverence that the summit of Haleakalā has always had within the indigenous Hawaiian community. We are most fortunate to have the opportunity to conduct observations from the mountain.

We thank the support of the staff at the Neil Gehrels Swift Observatory.

This work has made use of data from the Zwicky Transient Facility (ZTF). ZTF is supported by NSF grant No. AST- 1440341 and a collaboration including Caltech, IPAC, the Weizmann Institute for Science, the Oskar Klein Center at Stockholm University, the University of Maryland, the University of Washington, Deutsches Elektronen-Synchrotron and Humboldt University, Los Alamos National Laboratories, the TANGO Consortium of Taiwan, the University of Wisconsin–Milwaukee, and Lawrence Berkeley National Laboratories. Operations are conducted by COO, IPAC, and UW. The ZTF forced-photometry service was funded under the Heising-Simons Foundation grant No. 12540303 (PI: Graham).

This work has made use of data from the Asteroid Terrestrial-impact Last Alert System (ATLAS) project. The Asteroid Terrestrial-impact Last Alert System (ATLAS) project is primarily funded to search for near-earth asteroids through NASA grants NN12AR55G, 80NSSC18K0284, and 80NSSC18K1575; byproducts of the NEO search include images and catalogs from the survey area. This work was partially funded by Kepler/K2 grant J1944/80NSSC19K0112 and HST GO-15889, and STFC grants ST/T000198/1 and ST/S006109/1. The ATLAS science products have been made possible through the contributions of the University of Hawaii Institute for Astronomy, the Queen’s University Belfast, the Space Telescope Science Institute, the South African Astronomical Observatory, and The Millennium Institute of Astrophysics (MAS), Chile.

The PS1 and the PS1 public science archives have been made possible through contributions by the Institute for Astronomy, the University of Hawaii, the Pan-STARRS Project Office, the Max-Planck Society and its participating institutes, the Max Planck Institute for Astronomy, Heidelberg and the Max Planck Institute for Extraterrestrial Physics, Garching, Johns Hopkins University, Durham University, the University of Edinburgh, the Queen’s University Belfast, the Harvard-Smithsonian Center for Astrophysics, the Las Cumbres Observatory Global Telescope Network Incorporated, the National Central University of Taiwan, the Space Telescope Science Institute, NASA under grant No. NNX08AR22G issued through the Planetary Science Division of the NASA Science Mission Directorate, NSF grant No. AST- 1238877, the University of Maryland, Eotvos Lorand University, the Los Alamos National Laboratory, and the Gordon and Betty Moore Foundation.

This research has made use of the NASA Astrophysics Data System (ADS), the NASA/IPAC Extragalactic Database (NED), and NASA/IPAC Infrared Science Archive (IRSA, which is funded by NASA and operated by the California Institute of Technology) and IRAF (which is distributed by the National Optical Astronomy Observatory, NOAO, operated by the Association of Universities for Research in Astronomy, AURA, Inc., under cooperative agreement with the NSF).

TNS is supported by funding from the Weizmann Institute of Science, as well as grants from the Israeli Institute for Advanced Studies and the European Union via ERC grant No. 725161.

\end{acknowledgments}

\vspace{5mm}
\facilities{ATLAS, LCO, Swift(XRT and UVOT) and ZTF, MMT}

\software{astropy~\citep{2013A&A...558A..33A,2018AJ....156..123A}, \sw{SExtractor}~\citep{1996A&AS..117..393B} \sw{NumPy}~\citep{oliphant2015guide}, \sw{photutils}~\citep{2022zndo...6825092B}, \sw{PyRAF}~\citep{2012ascl.soft07011S}, \sw{SciPy}~\citep{2020SciPy-NMeth} and \sw{MOSFiT}~\citep{2017ascl.soft10006G}}

\appendix
\section{Photometry}

\begin{longtable}{|c|c|c|c|c|} 
\hline
\hline
MJD &  Filter &  Magnitude  $\pm$  e\_magnitude (AB) & Telescope \\
\hline
\hline
60170.40884 & r & $  > 20.55$  &   ZTF\\
60173.33883 & r & $  > 20.4$  &   ZTF\\
60173.47114 & g & $  > 20.47$  &   ZTF\\
60178.42973 & g & $  > 20.47$  &   ZTF\\
60180.40827 & g & $  > 19.18$  &   ZTF\\
60182.34443 & r & $  > 20.44$  &   ZTF\\
60182.42775 & g & $  > 20.28$  &   ZTF\\
60184.42790 & g & $  > 20.32$  &   ZTF\\
60184.47108 & r & $  > 20.46$  &   ZTF\\
60186.35961 & r & $  > 19.53$  &   ZTF\\
60186.47100 & g & $  > 19.43$  &   ZTF\\
60191.38920 & g & $  > 19.77$  &   ZTF\\
60194.30874 & g & $  > 20.32$  &   ZTF\\
60194.32521 & r & $  > 20.37$  &   ZTF\\
60196.32297 & g & $  20.68 \pm  0.33$  &   ZTF\\
60196.34751 & r & $  > 20.42$  &   ZTF\\
60200.37221 & g & $  20.30 \pm  0.20$  &   ZTF\\
60200.41501 & r & $  > 20.56$  &   ZTF\\
60202.26614 & r & $  20.40 \pm  0.23$  &   ZTF\\
60202.42600 & c &  $  19.79 \pm  0.21$  &   ATLAS\\
60202.43773 & g & $  20.28 \pm  0.24$  &   ZTF\\
60204.35063 & g & $  19.92 \pm  0.18$  &   ZTF\\
60204.40510 & r & $  20.39 \pm  0.19$  &   ZTF\\
60206.41362 & g & $  19.66 \pm  0.15$  &   ZTF\\
60206.42949 & r & $  20.16 \pm  0.18$  &   ZTF\\
60208.50777 & c &  $  19.77 \pm  0.14$  &   ATLAS\\
60209.31043 & r & $  19.74 \pm  0.17$  &   ZTF\\
60209.34994 & g & $  19.61 \pm  0.19$  &   ZTF\\
60210.49498 & c &  $  19.56 \pm  0.22$  &   ATLAS\\
60212.38339 & r & $  19.71 \pm  0.13$  &   ZTF\\
60212.41900 & g & $  19.40 \pm  0.13$  &   ZTF\\
60212.47191 & o & $  19.50 \pm  0.16$  &   ATLAS\\
60214.27741 & r & $  19.69 \pm  0.17$  &   ZTF\\
60214.36952 & g & $  19.24 \pm  0.18$  &   ZTF\\
60214.41461 & o & $  19.23 \pm  0.17$  &   ATLAS\\
60216.31725 & g & $  > 19.21$  &   ZTF\\
60216.48567 & r & $  19.35 \pm  0.24$  &   ZTF\\
60221.17881 & g & $  19.04 \pm  0.12$  &   ZTF\\
60221.28443 & r & $  19.47 \pm  0.19$  &   ZTF\\
60222.43430 & o & $  19.78 \pm  0.18$  &   ATLAS\\
60223.21542 & r & $  19.20 \pm  0.11$  &   ZTF\\
60223.28827 & g & $  18.88 \pm  0.11$  &   ZTF\\
60224.42465 & c &  $  19.06 \pm  0.09$  &   ATLAS\\
60226.28203 & r & $  19.12 \pm  0.11$  &   ZTF\\
60228.25706 & g & $  18.76 \pm  0.09$  &   ZTF\\
60228.40906 & c &  $  18.91 \pm  0.07$  &   ATLAS\\
60230.42111 & c &  $  19.10 \pm  0.07$  &   ATLAS\\
60231.19518 & g & $  18.69 \pm  0.08$  &   ZTF\\
60231.33180 & r & $  19.17 \pm  0.10$  &   ZTF\\
60233.17899 & r & $  19.10 \pm  0.10$  &   ZTF\\
60233.21849 & g & $  18.72 \pm  0.10$  &   ZTF\\
60234.42945 & c &  $  18.94 \pm  0.07$  &   ATLAS\\
60235.20121 & g & $  18.72 \pm  0.10$  &   ZTF\\
60235.24312 & r & $  19.07 \pm  0.10$  &   ZTF\\
60236.24132 & c &  $  18.93 \pm  0.08$  &   ATLAS\\
60237.19858 & r & $  19.14 \pm  0.11$  &   ZTF\\
60237.21539 & g & $  18.70 \pm  0.11$  &   ZTF\\
60238.38958 & c &  $  18.85 \pm  0.07$  &   ATLAS\\
60239.28438 & g & $  18.69 \pm  0.09$  &   ZTF\\
60239.34452 & r & $  19.07 \pm  0.09$  &   ZTF\\
60240.39917 & o & $  19.20 \pm  0.11$  &   ATLAS\\
60242.24405 & r & $  19.14 \pm  0.15$  &   ZTF\\
60242.31850 & g & $  18.63 \pm  0.10$  &   ZTF\\
60242.39736 & o & $  19.06 \pm  0.14$  &   ATLAS\\
60245.22380 & r & $  19.11 \pm  0.18$  &   ZTF\\
60245.28513 & g & $  18.61 \pm  0.13$  &   ZTF\\
60250.18170 & g & $  18.75 \pm  0.11$  &   ZTF\\
60250.26829 & r & $  19.21 \pm  0.12$  &   ZTF\\
60250.40169 & o & $  19.27 \pm  0.10$  &   ATLAS\\
60252.19480 & g & $  18.82 \pm  0.12$  &   ZTF\\
60252.28918 & r & $  19.06 \pm  0.10$  &   ZTF\\
60252.37470 & c &  $  19.10 \pm  0.09$  &   ATLAS\\
60254.14199 & r & $  19.15 \pm  0.10$  &   ZTF\\
60254.19508 & g & $  18.79 \pm  0.09$  &   ZTF\\
60254.36620 & c &  $  18.98 \pm  0.10$  &   ATLAS\\
60256.37487 & c &  $  19.09 \pm  0.10$  &   ATLAS\\
60257.17672 & r & $  19.31 \pm  0.14$  &   ZTF\\
60257.22870 & g & $  18.78 \pm  0.14$  &   ZTF\\
60258.36648 & c &  $  19.01 \pm  0.06$  &   ATLAS\\
60259.20232 & g & $  18.84 \pm  0.12$  &   ZTF\\
60259.29476 & r & $  19.32 \pm  0.14$  &   ZTF\\
60260.36655 & c &  $  19.26 \pm  0.12$  &   ATLAS\\
60261.15836 & r & $  19.32 \pm  0.10$  &   ZTF\\
60261.19790 & g & $  18.91 \pm  0.11$  &   ZTF\\
60263.17931 & g & $  > 18.28$  &   ZTF\\
60264.35645 & c &  $  19.22 \pm  0.11$  &   ATLAS\\
60266.34732 & c &  $  19.24 \pm  0.08$  &   ATLAS\\
60268.34586 & o & $  19.21 \pm  0.17$  &   ATLAS\\
60273.89381 & U & $  19.45 \pm  0.13$  &   LCO\\
60273.90333 & B & $  18.83 \pm  0.04$  &   LCO\\
60273.91468 & g & $  19.02 \pm  0.15$  &   LCO\\
60273.91851 & g & $  19.01 \pm  0.15$  &   LCO\\
60273.92247 & r & $  19.33 \pm  0.15$  &   LCO\\
60273.92513 & r & $  19.52 \pm  0.15$  &   LCO\\
60278.26281 & U & $  19.57 \pm  0.02$  &   LCO\\
60278.27241 & B & $  18.93 \pm  0.10$  &   LCO\\
60278.27964 & V & $  19.07 \pm  0.07$  &   LCO\\
60278.28382 & g & $  19.12 \pm  0.15$  &   LCO\\
60278.28763 & g & $  19.18 \pm  0.15$  &   LCO\\
60278.29163 & r & $  19.51 \pm  0.15$  &   LCO\\
60278.29428 & r & $  19.47 \pm  0.15$  &   LCO\\
60281.89540 & UVW1 & $  20.74 \pm  0.28$  &   Swift\\
60281.89610 & U & $  19.34 \pm  0.19$  &   Swift\\
60281.89646 & B & $  > 20.11$  &   Swift\\
60281.89735 & UVW2 & $  21.69 \pm  0.35$  &   Swift\\
60281.89839 & V & $  18.93 \pm  0.35$  &   Swift\\
60281.89965 & UVM2 & $  20.98 \pm  0.24$  &   Swift\\
60282.32095 & c &  $  19.14 \pm  0.11$  &   ATLAS\\
60283.24517 & U & $  19.59 \pm  0.07$  &   LCO\\
60283.25474 & B & $  19.06 \pm  0.02$  &   LCO\\
60283.26203 & V & $  19.20 \pm  0.01$  &   LCO\\
60283.26618 & g & $  19.17 \pm  0.15$  &   LCO\\
60283.27000 & g & $  19.16 \pm  0.15$  &   LCO\\
60283.27402 & r & $  19.47 \pm  0.15$  &   LCO\\
60283.27667 & r & $  19.50 \pm  0.15$  &   LCO\\
60283.33593 & o & $  19.54 \pm  0.13$  &   ATLAS\\
60284.15859 & r & $  19.60 \pm  0.14$  &   ZTF\\
60284.21976 & g & $  19.22 \pm  0.15$  &   ZTF\\
60284.26798 & o & $  19.51 \pm  0.16$  &   ATLAS\\
60284.29941 & c &  $  19.22 \pm  0.16$  &   ATLAS\\
60286.32714 & c &  $  19.49 \pm  0.13$  &   ATLAS\\
60287.50742 & UVW1 & $  20.07 \pm  0.23$  &   Swift\\
60287.50813 & U & $  19.53 \pm  0.26$  &   Swift\\
60287.50851 & B & $  > 19.89$  &   Swift\\
60287.50954 & UVW2 & $  > 20.3$  &   Swift\\
60287.51058 & V & $  > 19.03$  &   Swift\\
60287.51180 & UVM2 & $  20.52 \pm  0.24$  &   Swift\\
60288.13767 & g & $  19.31 \pm  0.20$  &   ZTF\\
60288.19723 & r & $  19.49 \pm  0.21$  &   ZTF\\
60288.90700 & U & $  19.62 \pm  0.17$  &   LCO\\
60288.91654 & B & $  19.17 \pm  0.03$  &   LCO\\
60288.92377 & V & $  19.17 \pm  0.14$  &   LCO\\
60288.92789 & g & $  19.23 \pm  0.15$  &   LCO\\
60288.93169 & g & $  19.25 \pm  0.15$  &   LCO\\
60288.93568 & r & $  19.77 \pm  0.15$  &   LCO\\
60288.93832 & r & $  19.79 \pm  0.15$  &   LCO\\
60289.35094 & V & $  > 18.62$  &   Swift\\
60289.35231 & UVM2 & $  20.13 \pm  0.26$  &   Swift\\
60289.41460 & UVW1 & $  20.02 \pm  0.22$  &   Swift\\
60289.41531 & U & $  19.28 \pm  0.23$  &   Swift\\
60289.41575 & B & $  > 19.74$  &   Swift\\
60289.41608 & UVW2 & $  20.81 \pm  0.33$  &   Swift\\
60290.18032 & r & $  19.78 \pm  0.17$  &   ZTF\\
60290.21987 & g & $  19.38 \pm  0.14$  &   ZTF\\
60291.34173 & o & $  19.44 \pm  0.15$  &   ATLAS\\
60291.37684 & c &  $  19.37 \pm  0.15$  &   ATLAS\\
60293.30967 & o & $  19.61 \pm  0.18$  &   ATLAS\\
60293.79962 & U & $  19.97 \pm  0.04$  &   LCO\\
60293.80916 & B & $  19.22 \pm  0.02$  &   LCO\\
60293.81639 & V & $  19.36 \pm  0.02$  &   LCO\\
60293.82054 & g & $  19.34 \pm  0.15$  &   LCO\\
60293.82436 & g & $  19.33 \pm  0.15$  &   LCO\\
60293.82836 & r & $  19.73 \pm  0.15$  &   LCO\\
60293.83103 & r & $  19.66 \pm  0.15$  &   LCO\\
60294.56479 & UVW1 & $  20.73 \pm  0.33$  &   Swift\\
60294.56645 & U & $  19.42 \pm  0.23$  &   Swift\\
60294.56730 & B & $  19.34 \pm  0.32$  &   Swift\\
60294.56976 & UVW2 & $  > 20.49$  &   Swift\\
60294.57222 & V & $  > 19.07$  &   Swift\\
60294.57557 & UVM2 & $  > 20.38$  &   Swift\\
60298.20710 & U & $  19.87 \pm  0.14$  &   LCO\\
60298.21665 & B & $  19.44 \pm  0.03$  &   LCO\\
60298.22388 & V & $  19.36 \pm  0.11$  &   LCO\\
60300.202080 & U & $  19.64 \pm  0.45$  &   LCO\\
60300.21166 & B & $  19.34 \pm  0.14$  &   LCO\\
60300.21891 & V & $  19.44 \pm  0.09$  &   LCO\\
60300.22308 & g & $  19.44 \pm  0.15$  &   LCO\\
60300.22691 & g & $  19.34 \pm  0.15$  &   LCO\\
60300.23089 & r & $  19.44 \pm  0.15$  &   LCO\\
60305.16276 & U & $  19.53 \pm  0.18$  &   LCO\\
60305.17233 & B & $  19.41 \pm  0.09$  &   LCO\\
60305.17959 & V & $  19.68 \pm  0.16$  &   LCO\\
60305.18374 & g & $  19.68 \pm  0.15$  &   LCO\\
60305.18755 & g & $  19.48 \pm  0.15$  &   LCO\\
60305.19154 & r & $  19.69 \pm  0.15$  &   LCO\\
60305.19421 & r & $  19.86 \pm  0.15$  &   LCO\\
60310.16442 & U & $  20.18 \pm  0.10$  &   LCO\\
60310.17395 & B & $  19.58 \pm  0.04$  &   LCO\\
60310.18117 & V & $  19.66 \pm  0.04$  &   LCO\\
60310.18531 & g & $  19.73 \pm  0.15$  &   LCO\\
60310.18913 & g & $  19.76 \pm  0.15$  &   LCO\\
60310.19314 & r & $  19.97 \pm  0.15$  &   LCO\\
60310.19578 & r & $  20.05 \pm  0.15$  &   LCO\\
60311.14469 & r & $  20.04 \pm  0.21$  &   ZTF\\
60315.10470 & U & $  20.26 \pm  0.08$  &   LCO\\
60315.11426 & B & $  19.78 \pm  0.02$  &   LCO\\
60315.12151 & V & $  19.66 \pm  0.07$  &   LCO\\
60315.12565 & g & $  19.84 \pm  0.15$  &   LCO\\
60315.12948 & g & $  19.84 \pm  0.15$  &   LCO\\
60315.13346 & r & $  20.17 \pm  0.15$  &   LCO\\
60315.13611 & r & $  19.85 \pm  0.15$  &   LCO\\
60317.10851 & U & $  19.85 \pm  0.05$  &   LCO\\
60317.11806 & B & $  19.71 \pm  0.07$  &   LCO\\
60317.12528 & V & $  19.63 \pm  0.03$  &   LCO\\
60317.12943 & g & $  20.01 \pm  0.15$  &   LCO\\
60317.13324 & g & $  19.93 \pm  0.15$  &   LCO\\
60317.13721 & r & $  20.18 \pm  0.15$  &   LCO\\
60318.91158 & U & $  20.41 \pm  0.02$  &   LCO\\
60318.92840 & V & $  19.71 \pm  0.02$  &   LCO\\
60318.93254 & g & $  19.84 \pm  0.15$  &   LCO\\
60318.93636 & g & $  19.88 \pm  0.15$  &   LCO\\
60318.94033 & r & $  20.04 \pm  0.15$  &   LCO\\
60318.94297 & r & $  20.06 \pm  0.15$  &   LCO\\
60323.89180 & U & $  20.51 \pm  0.11$  &   LCO\\
60323.90132 & B & $  19.81 \pm  0.06$  &   LCO\\
60323.90857 & V & $  19.79 \pm  0.10$  &   LCO\\
60323.91271 & g & $  19.97 \pm  0.15$  &   LCO\\
60323.91653 & g & $  20.03 \pm  0.15$  &   LCO\\
60323.92049 & r & $  20.17 \pm  0.15$  &   LCO\\
60323.92314 & r & $  20.29 \pm  0.15$  &   LCO\\
60328.87796 & U & $  20.48 \pm  0.15$  &   LCO\\
60328.88750 & B & $  19.94 \pm  0.07$  &   LCO\\
60328.89473 & V & $  19.77 \pm  0.08$  &   LCO\\
60328.89892 & g & $  20.13 \pm  0.15$  &   LCO\\
60328.90274 & g & $  20.01 \pm  0.15$  &   LCO\\
60328.90669 & r & $  20.30 \pm  0.15$  &   LCO\\
60328.90935 & r & $  20.25 \pm  0.15$  &   LCO\\
60337.09611 & U & $  20.78 \pm  0.03$  &   LCO\\
60337.10564 & B & $  19.91 \pm  0.05$  &   LCO\\
60337.11286 & V & $  19.52 \pm  0.17$  &   LCO\\
60337.12084 & g & $  20.07 \pm  0.15$  &   LCO\\
60337.12485 & r & $  20.11 \pm  0.15$  &   LCO\\
60337.12750 & r & $  20.47 \pm  0.15$  &   LCO\\
60342.87347 & g & $  20.33 \pm  0.15$  &   LCO\\
60342.87746 & r & $  20.49 \pm  0.15$  &   LCO\\
60347.86278 & g & $  20.40 \pm  0.15$  &   LCO\\
60347.86676 & r & $  20.62 \pm  0.15$  &   LCO\\
60352.84919 & g & $  20.46 \pm  0.15$  &   LCO\\
60352.85317 & r & $  20.51 \pm  0.15$  &   LCO\\
60357.08500 & g & $  20.39 \pm  0.15$  &   LCO\\
60357.08402 & r & $  20.80 \pm  0.15$  &   LCO\\

\hline

\caption{Photometry of AT2023vto. Magnitudes are corrected for Galactic Extinction in the direction of AT2023vto.}
\label{tab:photometrytable}
\end{longtable}

\bibliography{ref.bib}{}

\begin{thebibliography}{}
\expandafter\ifx\csname natexlab\endcsname\relax\def\natexlab#1{#1}\fi
\providecommand{\url}[1]{\href{#1}{#1}}
\providecommand{\dodoi}[1]{doi:~\href{http://doi.org/#1}{\nolinkurl{#1}}}
\providecommand{\doeprint}[1]{\href{http://ascl.net/#1}{\nolinkurl{http://ascl.net/#1}}}
\providecommand{\doarXiv}[1]{\href{https://arxiv.org/abs/#1}{\nolinkurl{https://arxiv.org/abs/#1}}}

\bibitem[{{Alexander} {et~al.}(2020){Alexander}, {van Velzen}, {Horesh}, \&
  {Zauderer}}]{Alexander2020}
{Alexander}, K.~D., {van Velzen}, S., {Horesh}, A., \& {Zauderer}, B.~A. 2020,
  \ssr, 216, 81, \dodoi{10.1007/s11214-020-00702-w}

\bibitem[{{Anderson} {et~al.}(2014){Anderson}, {Dessart}, {Gutierrez}, {Hamuy},
  {Morrell}, {Phillips}, {Folatelli}, {Stritzinger}, {Freedman},
  {Gonz{\'a}lez-Gait{\'a}n}, {McCarthy}, {Suntzeff}, \&
  {Thomas-Osip}}]{2014MNRAS.441..671A}
{Anderson}, J.~P., {Dessart}, L., {Gutierrez}, C.~P., {et~al.} 2014, \mnras,
  441, 671, \dodoi{10.1093/mnras/stu610}

\bibitem[{{Arcavi} {et~al.}(2020){Arcavi}, {Burke}, {Nyiha}, {Howell},
  {Hiramatsu}, {McCully}, {Pellegrino}, \& {Gonzalez}}]{2020TNSCR3228....1A}
{Arcavi}, I., {Burke}, J., {Nyiha}, I., {et~al.} 2020, Transient Name Server
  Classification Report, 2020-3228, 1

\bibitem[{{Arcavi} {et~al.}(2014){Arcavi}, {Gal-Yam}, {Sullivan}, {Pan},
  {Cenko}, {Horesh}, {Ofek}, {De Cia}, {Yan}, {Yang}, {Howell}, {Tal},
  {Kulkarni}, {Tendulkar}, {Tang}, {Xu}, {Sternberg}, {Cohen}, {Bloom},
  {Nugent}, {Kasliwal}, {Perley}, {Quimby}, {Miller}, {Theissen}, \&
  {Laher}}]{2014ApJ...793...38A}
{Arcavi}, I., {Gal-Yam}, A., {Sullivan}, M., {et~al.} 2014, \apj, 793, 38,
  \dodoi{10.1088/0004-637X/793/1/38}

\bibitem[{{Astropy Collaboration} {et~al.}(2013){Astropy Collaboration},
  {Robitaille}, {Tollerud}, {Greenfield}, {Droettboom}, {Bray}, {Aldcroft},
  {Davis}, {Ginsburg}, {Price-Whelan}, {Kerzendorf}, {Conley}, {Crighton},
  {Barbary}, {Muna}, {Ferguson}, {Grollier}, {Parikh}, {Nair}, {Unther},
  {Deil}, {Woillez}, {Conseil}, {Kramer}, {Turner}, {Singer}, {Fox}, {Weaver},
  {Zabalza}, {Edwards}, {Azalee Bostroem}, {Burke}, {Casey}, {Crawford},
  {Dencheva}, {Ely}, {Jenness}, {Labrie}, {Lim}, {Pierfederici}, {Pontzen},
  {Ptak}, {Refsdal}, {Servillat}, \& {Streicher}}]{2013A&A...558A..33A}
{Astropy Collaboration}, {Robitaille}, T.~P., {Tollerud}, E.~J., {et~al.} 2013,
  \aap, 558, A33, \dodoi{10.1051/0004-6361/201322068}

\bibitem[{{Astropy Collaboration} {et~al.}(2018){Astropy Collaboration},
  {Price-Whelan}, {Sip{\H{o}}cz}, {G{\"u}nther}, {Lim}, {Crawford}, {Conseil},
  {Shupe}, {Craig}, {Dencheva}, {Ginsburg}, {VanderPlas}, {Bradley},
  {P{\'e}rez-Su{\'a}rez}, {de Val-Borro}, {Aldcroft}, {Cruz}, {Robitaille},
  {Tollerud}, {Ardelean}, {Babej}, {Bach}, {Bachetti}, {Bakanov}, {Bamford},
  {Barentsen}, {Barmby}, {Baumbach}, {Berry}, {Biscani}, {Boquien}, {Bostroem},
  {Bouma}, {Brammer}, {Bray}, {Breytenbach}, {Buddelmeijer}, {Burke},
  {Calderone}, {Cano Rodr{\'\i}guez}, {Cara}, {Cardoso}, {Cheedella}, {Copin},
  {Corrales}, {Crichton}, {D'Avella}, {Deil}, {Depagne}, {Dietrich}, {Donath},
  {Droettboom}, {Earl}, {Erben}, {Fabbro}, {Ferreira}, {Finethy}, {Fox},
  {Garrison}, {Gibbons}, {Goldstein}, {Gommers}, {Greco}, {Greenfield},
  {Groener}, {Grollier}, {Hagen}, {Hirst}, {Homeier}, {Horton}, {Hosseinzadeh},
  {Hu}, {Hunkeler}, {Ivezi{\'c}}, {Jain}, {Jenness}, {Kanarek}, {Kendrew},
  {Kern}, {Kerzendorf}, {Khvalko}, {King}, {Kirkby}, {Kulkarni}, {Kumar},
  {Lee}, {Lenz}, {Littlefair}, {Ma}, {Macleod}, {Mastropietro}, {McCully},
  {Montagnac}, {Morris}, {Mueller}, {Mumford}, {Muna}, {Murphy}, {Nelson},
  {Nguyen}, {Ninan}, {N{\"o}the}, {Ogaz}, {Oh}, {Parejko}, {Parley}, {Pascual},
  {Patil}, {Patil}, {Plunkett}, {Prochaska}, {Rastogi}, {Reddy Janga},
  {Sabater}, {Sakurikar}, {Seifert}, {Sherbert}, {Sherwood-Taylor}, {Shih},
  {Sick}, {Silbiger}, {Singanamalla}, {Singer}, {Sladen}, {Sooley},
  {Sornarajah}, {Streicher}, {Teuben}, {Thomas}, {Tremblay}, {Turner},
  {Terr{\'o}n}, {van Kerkwijk}, {de la Vega}, {Watkins}, {Weaver}, {Whitmore},
  {Woillez}, {Zabalza}, \& {Astropy Contributors}}]{2018AJ....156..123A}
{Astropy Collaboration}, {Price-Whelan}, A.~M., {Sip{\H{o}}cz}, B.~M., {et~al.}
  2018, \aj, 156, 123, \dodoi{10.3847/1538-3881/aabc4f}

\bibitem[{{Auchettl} {et~al.}(2017){Auchettl}, {Guillochon}, \&
  {Ramirez-Ruiz}}]{2017ApJ...838..149A}
{Auchettl}, K., {Guillochon}, J., \& {Ramirez-Ruiz}, E. 2017, \apj, 838, 149,
  \dodoi{10.3847/1538-4357/aa633b}

\bibitem[{{Bellm}(2014)}]{2014htu..conf...27B}
{Bellm}, E. 2014, in The Third Hot-wiring the Transient Universe Workshop, ed.
  P.~R. {Wozniak}, M.~J. {Graham}, A.~A. {Mahabal}, \& R.~{Seaman}, 27--33,
  \dodoi{10.48550/arXiv.1410.8185}

\bibitem[{{Bellm} {et~al.}(2019){Bellm}, {Kulkarni}, {Graham}, {Dekany},
  {Smith}, {Riddle}, {Masci}, {Helou}, {Prince}, {Adams}, {Barbarino},
  {Barlow}, {Bauer}, {Beck}, {Belicki}, {Biswas}, {Blagorodnova}, {Bodewits},
  {Bolin}, {Brinnel}, {Brooke}, {Bue}, {Bulla}, {Burruss}, {Cenko}, {Chang},
  {Connolly}, {Coughlin}, {Cromer}, {Cunningham}, {De}, {Delacroix}, {Desai},
  {Duev}, {Eadie}, {Farnham}, {Feeney}, {Feindt}, {Flynn}, {Franckowiak},
  {Frederick}, {Fremling}, {Gal-Yam}, {Gezari}, {Giomi}, {Goldstein},
  {Golkhou}, {Goobar}, {Groom}, {Hacopians}, {Hale}, {Henning}, {Ho}, {Hover},
  {Howell}, {Hung}, {Huppenkothen}, {Imel}, {Ip}, {Ivezi{\'c}}, {Jackson},
  {Jones}, {Juric}, {Kasliwal}, {Kaspi}, {Kaye}, {Kelley}, {Kowalski},
  {Kramer}, {Kupfer}, {Landry}, {Laher}, {Lee}, {Lin}, {Lin}, {Lunnan},
  {Giomi}, {Mahabal}, {Mao}, {Miller}, {Monkewitz}, {Murphy}, {Ngeow},
  {Nordin}, {Nugent}, {Ofek}, {Patterson}, {Penprase}, {Porter}, {Rauch},
  {Rebbapragada}, {Reiley}, {Rigault}, {Rodriguez}, {van Roestel}, {Rusholme},
  {van Santen}, {Schulze}, {Shupe}, {Singer}, {Soumagnac}, {Stein}, {Surace},
  {Sollerman}, {Szkody}, {Taddia}, {Terek}, {Van Sistine}, {van Velzen},
  {Vestrand}, {Walters}, {Ward}, {Ye}, {Yu}, {Yan}, \&
  {Zolkower}}]{2019PASP..131a8002B}
{Bellm}, E.~C., {Kulkarni}, S.~R., {Graham}, M.~J., {et~al.} 2019, \pasp, 131,
  018002, \dodoi{10.1088/1538-3873/aaecbe}

\bibitem[{{Berger} {et~al.}(2012){Berger}, {Zauderer}, {Pooley}, {Soderberg},
  {Sari}, {Brunthaler}, \& {Bietenholz}}]{p1}
{Berger}, E., {Zauderer}, A., {Pooley}, G.~G., {et~al.} 2012, \apj, 748, 36,
  \dodoi{10.1088/0004-637X/748/1/36}

\bibitem[{{Bertin} \& {Arnouts}(1996)}]{1996A&AS..117..393B}
{Bertin}, E., \& {Arnouts}, S. 1996, \aaps, 117, 393,
  \dodoi{10.1051/aas:1996164}

\bibitem[{{Blagorodnova} {et~al.}(2017){Blagorodnova}, {Gezari}, {Hung},
  {Kulkarni}, {Cenko}, {Pasham}, {Yan}, {Arcavi}, {Ben-Ami}, {Bue}, {Cantwell},
  {Cao}, {Castro-Tirado}, {Fender}, {Fremling}, {Gal-Yam}, {Ho}, {Horesh},
  {Hosseinzadeh}, {Kasliwal}, {Kong}, {Laher}, {Leloudas}, {Lunnan}, {Masci},
  {Mooley}, {Neill}, {Nugent}, {Powell}, {Valeev}, {Vreeswijk}, {Walters}, \&
  {Wozniak}}]{2017ApJ...844...46B}
{Blagorodnova}, N., {Gezari}, S., {Hung}, T., {et~al.} 2017, \apj, 844, 46,
  \dodoi{10.3847/1538-4357/aa7579}

\bibitem[{{Blanchard} {et~al.}(2017){Blanchard}, {Nicholl}, {Berger},
  {Guillochon}, {Margutti}, {Chornock}, {Alexander}, {Leja}, \&
  {Drout}}]{2017ApJ...843..106B}
{Blanchard}, P.~K., {Nicholl}, M., {Berger}, E., {et~al.} 2017, \apj, 843, 106,
  \dodoi{10.3847/1538-4357/aa77f7}

\bibitem[{{Bradley} {et~al.}(2022){Bradley}, {Sip{\H{o}}cz}, {Robitaille},
  {Tollerud}, {Vin{\'\i}cius}, {Deil}, {Barbary}, {Wilson}, {Busko}, {Donath},
  {G{\"u}nther}, {Cara}, {Lim}, {Me{\ss}linger}, {Conseil}, {Bostroem},
  {Droettboom}, {Bray}, {Andersen Bratholm}, {Barentsen}, {Craig}, {Rathi},
  {Pascual}, {Perren}, {Georgiev}, {De Val-Borro}, {Kerzendorf}, {Bach},
  {Quint}, \& {Souchereau}}]{2022zndo...6825092B}
{Bradley}, L., {Sip{\H{o}}cz}, B., {Robitaille}, T., {et~al.} 2022,
  {astropy/photutils: 1.5.0}, 1.5.0,  Zenodo, \dodoi{10.5281/zenodo.6825092}

\bibitem[{{Brown} {et~al.}(2017){Brown}, {Levan}, {Stanway}, {Kr{\"u}hler},
  {Tanvir}, {Davies}, {Fruchter}, {Cenko}, \& {Metzger}}]{Brown2017}
{Brown}, G.~C., {Levan}, A.~J., {Stanway}, E.~R., {et~al.} 2017, \mnras, 472,
  4469, \dodoi{10.1093/mnras/stx2193}

\bibitem[{{Brown} {et~al.}(2014){Brown}, {Breeveld}, {Holland}, {Kuin}, \&
  {Pritchard}}]{2014Ap&SS.354...89B}
{Brown}, P.~J., {Breeveld}, A.~A., {Holland}, S., {Kuin}, P., \& {Pritchard},
  T. 2014, \apss, 354, 89, \dodoi{10.1007/s10509-014-2059-8}

\bibitem[{{Brown} {et~al.}(2009){Brown}, {Holland}, {Immler}, {Milne},
  {Roming}, {Gehrels}, {Nousek}, {Panagia}, {Still}, \& {Vanden
  Berk}}]{2009AJ....137.4517B}
{Brown}, P.~J., {Holland}, S.~T., {Immler}, S., {et~al.} 2009, \aj, 137, 4517,
  \dodoi{10.1088/0004-6256/137/5/4517}

\bibitem[{{Brown} {et~al.}(2016){Brown}, {Yang}, {Cooke}, {Olaes}, {Quimby},
  {Baade}, {Gehrels}, {Hoeflich}, {Maund}, {Mould}, {Wang}, \&
  {Wheeler}}]{2016ApJ...828....3B}
{Brown}, P.~J., {Yang}, Y., {Cooke}, J., {et~al.} 2016, \apj, 828, 3,
  \dodoi{10.3847/0004-637X/828/1/3}

\bibitem[{{Brown} {et~al.}(2013){Brown}, {Baliber}, {Bianco}, {Bowman},
  {Burleson}, {Conway}, {Crellin}, {Depagne}, {De Vera}, {Dilday}, {Dragomir},
  {Dubberley}, {Eastman}, {Elphick}, {Falarski}, {Foale}, {Ford}, {Fulton},
  {Garza}, {Gomez}, {Graham}, {Greene}, {Haldeman}, {Hawkins}, {Haworth},
  {Haynes}, {Hidas}, {Hjelstrom}, {Howell}, {Hygelund}, {Lister}, {Lobdill},
  {Martinez}, {Mullins}, {Norbury}, {Parrent}, {Paulson}, {Petry}, {Pickles},
  {Posner}, {Rosing}, {Ross}, {Sand}, {Saunders}, {Shobbrook}, {Shporer},
  {Street}, {Thomas}, {Tsapras}, {Tufts}, {Valenti}, {Vander Horst}, {Walker},
  {White}, \& {Willis}}]{2013PASP..125.1031B}
{Brown}, T.~M., {Baliber}, N., {Bianco}, F.~B., {et~al.} 2013, \pasp, 125,
  1031, \dodoi{10.1086/673168}

\bibitem[{{Cannizzaro} {et~al.}(2021){Cannizzaro}, {Wevers}, {Jonker},
  {P{\'e}rez-Torres}, {Moldon}, {Mata-S{\'a}nchez}, {Leloudas}, {Pasham},
  {Mattila}, {Arcavi}, {Decker French}, {Onori}, {Inserra}, {Nicholl},
  {Gromadzki}, {Chen}, {M{\"u}ller-Bravo}, {Short}, {Anderson}, {Young},
  {Gendreau}, {Arzoumanian}, {L{\"o}wenstein}, {Remillard}, {Roy}, \&
  {Hiramatsu}}]{2021MNRAS.504..792C}
{Cannizzaro}, G., {Wevers}, T., {Jonker}, P.~G., {et~al.} 2021, \mnras, 504,
  792, \dodoi{10.1093/mnras/stab851}

\bibitem[{{Cendes} {et~al.}(2023){Cendes}, {Berger}, {Alexander}, {Chornock},
  {Margutti}, {Metzger}, {Wieringa}, {Bietenholz}, {Hajela}, {Laskar}, {Stroh},
  \& {Terreran}}]{Cendes2023}
{Cendes}, Y., {Berger}, E., {Alexander}, K.~D., {et~al.} 2023, arXiv e-prints,
  arXiv:2308.13595, \dodoi{10.48550/arXiv.2308.13595}

\bibitem[{{Charalampopoulos} {et~al.}(2022){Charalampopoulos}, {Leloudas},
  {Malesani}, {Wevers}, {Arcavi}, {Nicholl}, {Pursiainen}, {Lawrence},
  {Anderson}, {Benetti}, {Cannizzaro}, {Chen}, {Galbany}, {Gromadzki},
  {Guti{\'e}rrez}, {Inserra}, {Jonker}, {M{\"u}ller-Bravo}, {Onori}, {Short},
  {Sollerman}, \& {Young}}]{2022A&A...659A..34C}
{Charalampopoulos}, P., {Leloudas}, G., {Malesani}, D.~B., {et~al.} 2022, \aap,
  659, A34, \dodoi{10.1051/0004-6361/202142122}

\bibitem[{{Chornock} {et~al.}(2014){Chornock}, {Berger}, {Gezari}, {Zauderer},
  {Rest}, {Chomiuk}, {Kamble}, {Soderberg}, {Czekala}, {Dittmann}, {Drout},
  {Foley}, {Fong}, {Huber}, {Kirshner}, {Lawrence}, {Lunnan}, {Marion},
  {Narayan}, {Riess}, {Roth}, {Sanders}, {Scolnic}, {Smartt}, {Smith},
  {Stubbs}, {Tonry}, {Burgett}, {Chambers}, {Flewelling}, {Hodapp}, {Kaiser},
  {Magnier}, {Martin}, {Neill}, {Price}, \& {Wainscoat}}]{2014ApJ...780...44C}
{Chornock}, R., {Berger}, E., {Gezari}, S., {et~al.} 2014, \apj, 780, 44,
  \dodoi{10.1088/0004-637X/780/1/44}

\bibitem[{{Dong} {et~al.}(2016){Dong}, {Shappee}, {Prieto}, {Jha}, {Stanek},
  {Holoien}, {Kochanek}, {Thompson}, {Morrell}, {Thompson}, {Basu}, {Beacom},
  {Bersier}, {Brimacombe}, {Brown}, {Bufano}, {Chen}, {Conseil}, {Danilet},
  {Falco}, {Grupe}, {Kiyota}, {Masi}, {Nicholls}, {Olivares E.}, {Pignata},
  {Pojmanski}, {Simonian}, {Szczygiel}, \& {Wo{\'z}niak}}]{2016Sci...351..257D}
{Dong}, S., {Shappee}, B.~J., {Prieto}, J.~L., {et~al.} 2016, Science, 351,
  257, \dodoi{10.1126/science.aac9613}

\bibitem[{{Drake} {et~al.}(2011){Drake}, {Djorgovski}, {Mahabal}, {Anderson},
  {Roy}, {Mohan}, {Ravindranath}, {Frail}, {Gezari}, {Neill}, {Ho}, {Prieto},
  {Thompson}, {Thorstensen}, {Wagner}, {Kowalski}, {Chiang}, {Grove},
  {Schinzel}, {Wood}, {Carrasco}, {Recillas}, {Kewley}, {Archana}, {Basu},
  {Wadadekar}, {Kumar}, {Myers}, {Phinney}, {Williams}, {Graham}, {Catelan},
  {Beshore}, {Larson}, \& {Christensen}}]{2011ApJ...735..106D}
{Drake}, A.~J., {Djorgovski}, S.~G., {Mahabal}, A., {et~al.} 2011, \apj, 735,
  106, \dodoi{10.1088/0004-637X/735/2/106}

\bibitem[{{Evans} \& {Kochanek}(1989)}]{1989ApJ...346L..13E}
{Evans}, C.~R., \& {Kochanek}, C.~S. 1989, \apjl, 346, L13,
  \dodoi{10.1086/185567}

\bibitem[{{Evans} {et~al.}(2007){Evans}, {Beardmore}, {Page}, {Tyler},
  {Osborne}, {Goad}, {O'Brien}, {Vetere}, {Racusin}, {Morris}, {Burrows},
  {Capalbi}, {Perri}, {Gehrels}, \& {Romano}}]{2007A&A...469..379E}
{Evans}, P.~A., {Beardmore}, A.~P., {Page}, K.~L., {et~al.} 2007, \aap, 469,
  379, \dodoi{10.1051/0004-6361:20077530}

\bibitem[{{Evans} {et~al.}(2009){Evans}, {Beardmore}, {Page}, {Osborne},
  {O'Brien}, {Willingale}, {Starling}, {Burrows}, {Godet}, {Vetere}, {Racusin},
  {Goad}, {Wiersema}, {Angelini}, {Capalbi}, {Chincarini}, {Gehrels}, {Kennea},
  {Margutti}, {Morris}, {Mountford}, {Pagani}, {Perri}, {Romano}, \&
  {Tanvir}}]{2009MNRAS.397.1177E}
---. 2009, \mnras, 397, 1177, \dodoi{10.1111/j.1365-2966.2009.14913.x}

\bibitem[{{Fabricant} {et~al.}(2019){Fabricant}, {Fata}, {Epps}, {Gauron},
  {Mueller}, {Zajac}, {Amato}, {Barberis}, {Bergner}, {Brennan}, {Brown},
  {Chilingarian}, {Geary}, {Kradinov}, {McLeod}, {Smith}, \&
  {Woods}}]{2019PASP..131g5004F}
{Fabricant}, D., {Fata}, R., {Epps}, H., {et~al.} 2019, \pasp, 131, 075004,
  \dodoi{10.1088/1538-3873/ab1d78}

\bibitem[{{Fitzpatrick}(1999)}]{1999PASP..111...63F}
{Fitzpatrick}, E.~L. 1999, \pasp, 111, 63, \dodoi{10.1086/316293}

\bibitem[{{Foreman-Mackey} {et~al.}(2013){Foreman-Mackey}, {Hogg}, {Lang}, \&
  {Goodman}}]{2013PASP..125..306F}
{Foreman-Mackey}, D., {Hogg}, D.~W., {Lang}, D., \& {Goodman}, J. 2013, \pasp,
  125, 306, \dodoi{10.1086/670067}

\bibitem[{{F{\"o}rster} {et~al.}(2021){F{\"o}rster}, {Cabrera-Vives},
  {Castillo-Navarrete}, {Est{\'e}vez}, {S{\'a}nchez-S{\'a}ez}, {Arredondo},
  {Bauer}, {Carrasco-Davis}, {Catelan}, {Elorrieta}, {Eyheramendy}, {Huijse},
  {Pignata}, {Reyes}, {Reyes}, {Rodr{\'\i}guez-Mancini}, {Ruz-Mieres},
  {Valenzuela}, {{\'A}lvarez-Maldonado}, {Astorga}, {Borissova}, {Clocchiatti},
  {De Cicco}, {Donoso-Oliva}, {Hern{\'a}ndez-Garc{\'\i}a}, {Graham},
  {Jord{\'a}n}, {Kurtev}, {Mahabal}, {Maureira}, {Mu{\~n}oz-Arancibia},
  {Molina-Ferreiro}, {Moya}, {Palma}, {P{\'e}rez-Carrasco}, {Protopapas},
  {Romero}, {Sabatini-Gacitua}, {S{\'a}nchez}, {San Mart{\'\i}n},
  {Sep{\'u}lveda-Cobo}, {Vera}, \& {Vergara}}]{2021AJ....161..242F}
{F{\"o}rster}, F., {Cabrera-Vives}, G., {Castillo-Navarrete}, E., {et~al.}
  2021, \aj, 161, 242, \dodoi{10.3847/1538-3881/abe9bc}

\bibitem[{{Fremling}(2021)}]{2021TNSTR3208....1F}
{Fremling}, C. 2021, Transient Name Server Discovery Report, 2021-3208, 1

\bibitem[{{Fremling}(2023)}]{2023TNSTR2714....1F}
---. 2023, Transient Name Server Discovery Report, 2023-2714, 1

\bibitem[{{Gehrels} {et~al.}(2004){Gehrels}, {Chincarini}, {Giommi}, {Mason},
  {Nousek}, {Wells}, {White}, {Barthelmy}, {Burrows}, {Cominsky}, {Hurley},
  {Marshall}, {M{\'e}sz{\'a}ros}, {Roming}, {Angelini}, {Barbier}, {Belloni},
  {Campana}, {Caraveo}, {Chester}, {Citterio}, {Cline}, {Cropper}, {Cummings},
  {Dean}, {Feigelson}, {Fenimore}, {Frail}, {Fruchter}, {Garmire}, {Gendreau},
  {Ghisellini}, {Greiner}, {Hill}, {Hunsberger}, {Krimm}, {Kulkarni}, {Kumar},
  {Lebrun}, {Lloyd-Ronning}, {Markwardt}, {Mattson}, {Mushotzky}, {Norris},
  {Osborne}, {Paczynski}, {Palmer}, {Park}, {Parsons}, {Paul}, {Rees},
  {Reynolds}, {Rhoads}, {Sasseen}, {Schaefer}, {Short}, {Smale}, {Smith},
  {Stella}, {Tagliaferri}, {Takahashi}, {Tashiro}, {Townsley}, {Tueller},
  {Turner}, {Vietri}, {Voges}, {Ward}, {Willingale}, {Zerbi}, \&
  {Zhang}}]{2004ApJ...611.1005G}
{Gehrels}, N., {Chincarini}, G., {Giommi}, P., {et~al.} 2004, \apj, 611, 1005,
  \dodoi{10.1086/422091}

\bibitem[{{Gezari}(2021)}]{2021ARA&A..59...21G}
{Gezari}, S. 2021, \araa, 59, 21, \dodoi{10.1146/annurev-astro-111720-030029}

\bibitem[{{Gezari} {et~al.}(2006){Gezari}, {Martin}, {Milliard}, {Basa},
  {Halpern}, {Forster}, {Friedman}, {Morrissey}, {Neff}, {Schiminovich},
  {Seibert}, {Small}, \& {Wyder}}]{2006ApJ...653L..25G}
{Gezari}, S., {Martin}, D.~C., {Milliard}, B., {et~al.} 2006, \apjl, 653, L25,
  \dodoi{10.1086/509918}

\bibitem[{{Gomez} {et~al.}(2023){Gomez}, {Villar}, {Berger}, {Gezari}, {van
  Velzen}, {Nicholl}, {Blanchard}, \& {Alexander}}]{2023ApJ...949..113G}
{Gomez}, S., {Villar}, V.~A., {Berger}, E., {et~al.} 2023, \apj, 949, 113,
  \dodoi{10.3847/1538-4357/acc535}

\bibitem[{{Gomez} {et~al.}(2020){Gomez}, {Nicholl}, {Short}, {Margutti},
  {Alexander}, {Blanchard}, {Berger}, {Eftekhari}, {Schulze}, {Anderson},
  {Arcavi}, {Chornock}, {Cowperthwaite}, {Galbany}, {Herzog}, {Hiramatsu},
  {Hosseinzadeh}, {Laskar}, {M{\"u}ller Bravo}, {Patton}, \&
  {Terreran}}]{2020MNRAS.497.1925G}
{Gomez}, S., {Nicholl}, M., {Short}, P., {et~al.} 2020, \mnras, 497, 1925,
  \dodoi{10.1093/mnras/staa2099}

\bibitem[{{Goodwin} {et~al.}(2023){Goodwin}, {Alexander}, {Miller-Jones},
  {Bietenholz}, {van Velzen}, {Anderson}, {Berger}, {Cendes}, {Chornock},
  {Coppejans}, {Eftekhari}, {Gezari}, {Laskar}, {Ramirez-Ruiz}, \&
  {Saxton}}]{2023MNRAS.522.5084G}
{Goodwin}, A.~J., {Alexander}, K.~D., {Miller-Jones}, J.~C.~A., {et~al.} 2023,
  \mnras, 522, 5084, \dodoi{10.1093/mnras/stad1258}

\bibitem[{{Greene} {et~al.}(2020){Greene}, {Strader}, \&
  {Ho}}]{2020ARA&A..58..257G}
{Greene}, J.~E., {Strader}, J., \& {Ho}, L.~C. 2020, \araa, 58, 257,
  \dodoi{10.1146/annurev-astro-032620-021835}

\bibitem[{{Guillochon} {et~al.}(2017){Guillochon}, {Nicholl}, {Villar},
  {Mockler}, {Narayan}, {Mandel}, {Berger}, \&
  {Williams}}]{2017ascl.soft10006G}
{Guillochon}, J., {Nicholl}, M., {Villar}, V.~A., {et~al.} 2017, {MOSFiT:
  Modular Open-Source Fitter for Transients}, Astrophysics Source Code Library,
  record ascl:1710.006

\bibitem[{{Guillochon} \& {Ramirez-Ruiz}(2015)}]{2015ApJ...798...64G}
{Guillochon}, J., \& {Ramirez-Ruiz}, E. 2015, \apj, 798, 64,
  \dodoi{10.1088/0004-637X/798/1/64}

\bibitem[{{Hammerstein} {et~al.}(2023){Hammerstein}, {van Velzen}, {Gezari},
  {Cenko}, {Yao}, {Ward}, {Frederick}, {Villanueva}, {Somalwar}, {Graham},
  {Kulkarni}, {Stern}, {Andreoni}, {Bellm}, {Dekany}, {Dhawan}, {Drake},
  {Fremling}, {Gatkine}, {Groom}, {Ho}, {Kasliwal}, {Karambelkar}, {Kool},
  {Masci}, {Medford}, {Perley}, {Purdum}, {van Roestel}, {Sharma}, {Sollerman},
  {Taggart}, \& {Yan}}]{2023ApJ...942....9H}
{Hammerstein}, E., {van Velzen}, S., {Gezari}, S., {et~al.} 2023, \apj, 942, 9,
  \dodoi{10.3847/1538-4357/aca283}

\bibitem[{{Henden} {et~al.}(2016){Henden}, {Templeton}, {Terrell}, {Smith},
  {Levine}, \& {Welch}}]{2016yCat.2336....0H}
{Henden}, A.~A., {Templeton}, M., {Terrell}, D., {et~al.} 2016, VizieR Online
  Data Catalog, II/336

\bibitem[{{HI4PI Collaboration} {et~al.}(2016){HI4PI Collaboration}, {Ben
  Bekhti}, {Fl{\"o}er}, {Keller}, {Kerp}, {Lenz}, {Winkel}, {Bailin},
  {Calabretta}, {Dedes}, {Ford}, {Gibson}, {Haud}, {Janowiecki}, {Kalberla},
  {Lockman}, {McClure-Griffiths}, {Murphy}, {Nakanishi}, {Pisano}, \&
  {Staveley-Smith}}]{2016A&A...594A.116H}
{HI4PI Collaboration}, {Ben Bekhti}, N., {Fl{\"o}er}, L., {et~al.} 2016, \aap,
  594, A116, \dodoi{10.1051/0004-6361/201629178}

\bibitem[{{Hinkle}(2024)}]{2024MNRAS.531.2603H}
{Hinkle}, J.~T. 2024, \mnras, 531, 2603, \dodoi{10.1093/mnras/stae1229}

\bibitem[{{Hinkle} {et~al.}(2024){Hinkle}, {Shappee}, {Auchettl}, {Kochanek},
  {Neustadt}, {Polin}, {Strader}, {Holoien}, {Huber}, {Tucker}, {Ashall}, {de
  Jaeger}, {Desai}, {Do}, {Hoogendam}, \& {Payne}}]{2024arXiv240508855H}
{Hinkle}, J.~T., {Shappee}, B.~J., {Auchettl}, K., {et~al.} 2024, arXiv
  e-prints, arXiv:2405.08855, \dodoi{10.48550/arXiv.2405.08855}

\bibitem[{{Hiramatsu} {et~al.}(2024){Hiramatsu}, {Matsumoto}, {Berger},
  {Ransome}, {Villar}, {Gomez}, {Cendes}, {De}, {Bostroem}, {Farah}, {Howell},
  {McCully}, {Newsome}, {Padilla Gonzalez}, {Pellegrino}, {Suzuki}, \&
  {Terreran}}]{2024ApJ...964..181H}
{Hiramatsu}, D., {Matsumoto}, T., {Berger}, E., {et~al.} 2024, \apj, 964, 181,
  \dodoi{10.3847/1538-4357/ad2854}

\bibitem[{{Holoien} {et~al.}(2014){Holoien}, {Prieto}, {Bersier}, {Kochanek},
  {Stanek}, {Shappee}, {Grupe}, {Basu}, {Beacom}, {Brimacombe}, {Brown},
  {Davis}, {Jencson}, {Pojmanski}, \& {Szczygie{\l}}}]{2014MNRAS.445.3263H}
{Holoien}, T.~W.~S., {Prieto}, J.~L., {Bersier}, D., {et~al.} 2014, \mnras,
  445, 3263, \dodoi{10.1093/mnras/stu1922}

\bibitem[{{Holoien} {et~al.}(2016{\natexlab{a}}){Holoien}, {Kochanek},
  {Prieto}, {Stanek}, {Dong}, {Shappee}, {Grupe}, {Brown}, {Basu}, {Beacom},
  {Bersier}, {Brimacombe}, {Danilet}, {Falco}, {Guo}, {Jose}, {Herczeg},
  {Long}, {Pojmanski}, {Simonian}, {Szczygie{\l}}, {Thompson}, {Thorstensen},
  {Wagner}, \& {Wo{\'z}niak}}]{2016MNRAS.455.2918H}
{Holoien}, T.~W.~S., {Kochanek}, C.~S., {Prieto}, J.~L., {et~al.}
  2016{\natexlab{a}}, \mnras, 455, 2918, \dodoi{10.1093/mnras/stv2486}

\bibitem[{{Holoien} {et~al.}(2016{\natexlab{b}}){Holoien}, {Kochanek},
  {Prieto}, {Grupe}, {Chen}, {Godoy-Rivera}, {Stanek}, {Shappee}, {Dong},
  {Brown}, {Basu}, {Beacom}, {Bersier}, {Brimacombe}, {Carlson}, {Falco},
  {Johnston}, {Madore}, {Pojmanski}, \& {Seibert}}]{2016MNRAS.463.3813H}
---. 2016{\natexlab{b}}, \mnras, 463, 3813, \dodoi{10.1093/mnras/stw2272}

\bibitem[{{Holoien} {et~al.}(2019){Holoien}, {Vallely}, {Auchettl}, {Stanek},
  {Kochanek}, {French}, {Prieto}, {Shappee}, {Brown}, {Fausnaugh}, {Dong},
  {Thompson}, {Bose}, {Neustadt}, {Cacella}, {Brimacombe}, {Kendurkar},
  {Beaton}, {Boutsia}, {Chomiuk}, {Connor}, {Morrell}, {Newman}, {Rudie},
  {Shishkovksy}, \& {Strader}}]{2019ApJ...883..111H}
{Holoien}, T. W.~S., {Vallely}, P.~J., {Auchettl}, K., {et~al.} 2019, \apj,
  883, 111, \dodoi{10.3847/1538-4357/ab3c66}

\bibitem[{{Holoien} {et~al.}(2022){Holoien}, {Neustadt}, {Vallely}, {Auchettl},
  {Hinkle}, {Romero-Ca{\~n}izales}, {Shappee}, {Kochanek}, {Stanek}, {Chen},
  {Dong}, {Prieto}, {Thompson}, {Brink}, {Filippenko}, {Zheng}, {Bersier},
  {Bose}, {Burgasser}, {Channa}, {de Jaeger}, {Hestenes}, {Im}, {Jeffers},
  {Jun}, {Lansbury}, {Post}, {Ross}, {Stern}, {Tang}, {Tucker}, {Valenti},
  {Yunus}, \& {Zhang}}]{2022ApJ...933..196H}
{Holoien}, T. W.~S., {Neustadt}, J. M.~M., {Vallely}, P.~J., {et~al.} 2022,
  \apj, 933, 196, \dodoi{10.3847/1538-4357/ac74b9}

\bibitem[{{Howell} \& {Global Supernova Project}(2017)}]{2017AAS...23031803H}
{Howell}, D.~A., \& {Global Supernova Project}. 2017, in American Astronomical
  Society Meeting Abstracts, Vol. 230, American Astronomical Society Meeting
  Abstracts \#230, 318.03

\bibitem[{{Hung} {et~al.}(2017){Hung}, {Gezari}, {Blagorodnova}, {Roth},
  {Cenko}, {Kulkarni}, {Horesh}, {Arcavi}, {McCully}, {Yan}, {Lunnan},
  {Fremling}, {Cao}, {Nugent}, \& {Wozniak}}]{2017ApJ...842...29H}
{Hung}, T., {Gezari}, S., {Blagorodnova}, N., {et~al.} 2017, \apj, 842, 29,
  \dodoi{10.3847/1538-4357/aa7337}

\bibitem[{{Hung} {et~al.}(2019){Hung}, {Cenko}, {Roth}, {Gezari}, {Veilleux},
  {van Velzen}, {Gaskell}, {Foley}, {Blagorodnova}, {Yan}, {Graham}, {Brown},
  {Siebert}, {Frederick}, {Ward}, {Gatkine}, {Gal-Yam}, {Yang}, {Schulze},
  {Dimitriadis}, {Kupfer}, {Shupe}, {Rusholme}, {Masci}, {Riddle}, {Soumagnac},
  {van Roestel}, \& {Dekany}}]{2019ApJ...879..119H}
{Hung}, T., {Cenko}, S.~B., {Roth}, N., {et~al.} 2019, \apj, 879, 119,
  \dodoi{10.3847/1538-4357/ab24de}

\bibitem[{{Hung} {et~al.}(2021){Hung}, {Foley}, {Veilleux}, {Cenko}, {Dai},
  {Auchettl}, {Brink}, {Dimitriadis}, {Filippenko}, {Gezari}, {Holoien},
  {Kilpatrick}, {Mockler}, {Piro}, {Ramirez-Ruiz}, {Rojas-Bravo}, {Siebert},
  {van Velzen}, \& {Zheng}}]{2021ApJ...917....9H}
{Hung}, T., {Foley}, R.~J., {Veilleux}, S., {et~al.} 2021, \apj, 917, 9,
  \dodoi{10.3847/1538-4357/abf4c3}

\bibitem[{{Inserra} {et~al.}(2018){Inserra}, {Smartt}, {Gall}, {Leloudas},
  {Chen}, {Schulze}, {Jerkstrand}, {Nicholl}, {Anderson}, {Arcavi}, {Benetti},
  {Cartier}, {Childress}, {Della Valle}, {Flewelling}, {Fraser}, {Gal-Yam},
  {Guti{\'e}rrez}, {Hosseinzadeh}, {Howell}, {Huber}, {Kankare}, {Kr{\"u}hler},
  {Magnier}, {Maguire}, {McCully}, {Prajs}, {Primak}, {Scalzo}, {Schmidt},
  {Smith}, {Smith}, {Tucker}, {Valenti}, {Wilman}, {Young}, \&
  {Yuan}}]{2018MNRAS.475.1046I}
{Inserra}, C., {Smartt}, S.~J., {Gall}, E.~E.~E., {et~al.} 2018, \mnras, 475,
  1046, \dodoi{10.1093/mnras/stx3179}

\bibitem[{{Kankare} {et~al.}(2017){Kankare}, {Kotak}, {Mattila}, {Lundqvist},
  {Ward}, {Fraser}, {Lawrence}, {Smartt}, {Meikle}, {Bruce}, {Harmanen},
  {Hutton}, {Inserra}, {Kangas}, {Pastorello}, {Reynolds},
  {Romero-Ca{\~n}izales}, {Smith}, {Valenti}, {Chambers}, {Hodapp}, {Huber},
  {Kaiser}, {Kudritzki}, {Magnier}, {Tonry}, {Wainscoat}, \&
  {Waters}}]{2017NatAs...1..865K}
{Kankare}, E., {Kotak}, R., {Mattila}, S., {et~al.} 2017, Nature Astronomy, 1,
  865, \dodoi{10.1038/s41550-017-0290-2}

\bibitem[{{Kippenhahn} {et~al.}(2013){Kippenhahn}, {Weigert}, \&
  {Weiss}}]{2013sse..book.....K}
{Kippenhahn}, R., {Weigert}, A., \& {Weiss}, A. 2013, {Stellar Structure and
  Evolution}, \dodoi{10.1007/978-3-642-30304-3}

\bibitem[{{Leja} {et~al.}(2017){Leja}, {Johnson}, {Conroy}, {van Dokkum}, \&
  {Byler}}]{2017ApJ...837..170L}
{Leja}, J., {Johnson}, B.~D., {Conroy}, C., {van Dokkum}, P.~G., \& {Byler}, N.
  2017, \apj, 837, 170, \dodoi{10.3847/1538-4357/aa5ffe}

\bibitem[{{Leloudas} {et~al.}(2016){Leloudas}, {Fraser}, {Stone}, {van Velzen},
  {Jonker}, {Arcavi}, {Fremling}, {Maund}, {Smartt}, {Kr{\`\i}hler},
  {Miller-Jones}, {Vreeswijk}, {Gal-Yam}, {Mazzali}, {De Cia}, {Howell},
  {Inserra}, {Patat}, {de Ugarte Postigo}, {Yaron}, {Ashall}, {Bar},
  {Campbell}, {Chen}, {Childress}, {Elias-Rosa}, {Harmanen}, {Hosseinzadeh},
  {Johansson}, {Kangas}, {Kankare}, {Kim}, {Kuncarayakti}, {Lyman}, {Magee},
  {Maguire}, {Malesani}, {Mattila}, {McCully}, {Nicholl}, {Prentice},
  {Romero-Ca{\~n}izales}, {Schulze}, {Smith}, {Sollerman}, {Sullivan},
  {Tucker}, {Valenti}, {Wheeler}, \& {Young}}]{2016NatAs...1E...2L}
{Leloudas}, G., {Fraser}, M., {Stone}, N.~C., {et~al.} 2016, Nature Astronomy,
  1, 0002, \dodoi{10.1038/s41550-016-0002}

\bibitem[{{Liu} {et~al.}(2022){Liu}, {Dou}, {Chen}, \&
  {Shen}}]{2022ApJ...925...67L}
{Liu}, X.-L., {Dou}, L.-M., {Chen}, J.-H., \& {Shen}, R.-F. 2022, \apj, 925,
  67, \dodoi{10.3847/1538-4357/ac33a9}

\bibitem[{{McMullin} {et~al.}(2007){McMullin}, {Waters}, {Schiebel}, {Young},
  \& {Golap}}]{McMullin2007}
{McMullin}, J.~P., {Waters}, B., {Schiebel}, D., {Young}, W., \& {Golap}, K.
  2007, in Astronomical Society of the Pacific Conference Series, Vol. 376,
  Astronomical Data Analysis Software and Systems XVI, ed. R.~A. {Shaw},
  F.~{Hill}, \& D.~J. {Bell}, 127

\bibitem[{{Mockler} {et~al.}(2019){Mockler}, {Guillochon}, \&
  {Ramirez-Ruiz}}]{2019ApJ...872..151M}
{Mockler}, B., {Guillochon}, J., \& {Ramirez-Ruiz}, E. 2019, \apj, 872, 151,
  \dodoi{10.3847/1538-4357/ab010f}

\bibitem[{{Neustadt} {et~al.}(2020){Neustadt}, {Holoien}, {Kochanek},
  {Auchettl}, {Brown}, {Shappee}, {Pogge}, {Dong}, {Stanek}, {Tucker}, {Bose},
  {Chen}, {Ricci}, {Vallely}, {Prieto}, {Thompson}, {Coulter}, {Drout},
  {Foley}, {Kilpatrick}, {Piro}, {Rojas-Bravo}, {Buckley}, {Gromadzki},
  {Dimitriadis}, {Siebert}, {Do}, {Huber}, \& {Payne}}]{2020MNRAS.494.2538N}
{Neustadt}, J.~M.~M., {Holoien}, T.~W.~S., {Kochanek}, C.~S., {et~al.} 2020,
  \mnras, 494, 2538, \dodoi{10.1093/mnras/staa859}

\bibitem[{{Nicholl} {et~al.}(2022){Nicholl}, {Lanning}, {Ramsden}, {Mockler},
  {Lawrence}, {Short}, \& {Ridley}}]{2022MNRAS.515.5604N}
{Nicholl}, M., {Lanning}, D., {Ramsden}, P., {et~al.} 2022, \mnras, 515, 5604,
  \dodoi{10.1093/mnras/stac2206}

\bibitem[{{Nicholl} {et~al.}(2019){Nicholl}, {Blanchard}, {Berger}, {Gomez},
  {Margutti}, {Alexander}, {Guillochon}, {Leja}, {Chornock}, {Snios},
  {Auchettl}, {Bruce}, {Challis}, {D'Orazio}, {Drout}, {Eftekhari}, {Foley},
  {Graur}, {Kilpatrick}, {Lawrence}, {Piro}, {Rojas-Bravo}, {Ross}, {Short},
  {Smartt}, {Smith}, \& {Stalder}}]{2019MNRAS.488.1878N}
{Nicholl}, M., {Blanchard}, P.~K., {Berger}, E., {et~al.} 2019, \mnras, 488,
  1878, \dodoi{10.1093/mnras/stz1837}

\bibitem[{{Nicholl} {et~al.}(2020){Nicholl}, {Wevers}, {Oates}, {Alexander},
  {Leloudas}, {Onori}, {Jerkstrand}, {Gomez}, {Campana}, {Arcavi},
  {Charalampopoulos}, {Gromadzki}, {Ihanec}, {Jonker}, {Lawrence}, {Mandel},
  {Schulze}, {Short}, {Burke}, {McCully}, {Hiramatsu}, {Howell}, {Pellegrino},
  {Abbot}, {Anderson}, {Berger}, {Blanchard}, {Cannizzaro}, {Chen},
  {Dennefeld}, {Galbany}, {Gonz{\'a}lez-Gait{\'a}n}, {Hosseinzadeh}, {Inserra},
  {Irani}, {Kuin}, {M{\"u}ller-Bravo}, {Pineda}, {Ross}, {Roy}, {Smartt},
  {Smith}, {Tucker}, {Wyrzykowski}, \& {Young}}]{2020MNRAS.499..482N}
{Nicholl}, M., {Wevers}, T., {Oates}, S.~R., {et~al.} 2020, \mnras, 499, 482,
  \dodoi{10.1093/mnras/staa2824}

\bibitem[{{Nordin} {et~al.}(2020){Nordin}, {Brinnel}, {Giomi}, {Santen},
  {Gal-Yam}, {Yaron}, \& {Schulze}}]{2020TNSTR3332....1N}
{Nordin}, J., {Brinnel}, V., {Giomi}, M., {et~al.} 2020, Transient Name Server
  Discovery Report, 2020-3332, 1

\bibitem[{Oliphant(2015)}]{oliphant2015guide}
Oliphant, T.~E. 2015, USA: CreateS-pace Independent Publishing Platform

\bibitem[{{Parkinson} {et~al.}(2022){Parkinson}, {Knigge}, {Matthews}, {Long},
  {Higginbottom}, {Sim}, \& {Mangham}}]{2022MNRAS.510.5426P}
{Parkinson}, E.~J., {Knigge}, C., {Matthews}, J.~H., {et~al.} 2022, \mnras,
  510, 5426, \dodoi{10.1093/mnras/stac027}

\bibitem[{{Perez-Fournon} {et~al.}(2020){Perez-Fournon}, {Angel}, {Poidevin},
  {Shirley}, {Marques-Chaves}, {Geier}, {Shu}, {Rodney}, {Roberts-Pierel},
  {Bolton}, {Chakrabarti}, {Craig}, \& {Alamiri}}]{2020TNSTR2500....1P}
{Perez-Fournon}, I., {Angel}, C.~J., {Poidevin}, F., {et~al.} 2020, Transient
  Name Server Discovery Report, 2020-2500, 1

\bibitem[{{Poidevin} {et~al.}(2023){Poidevin}, {P{\'e}rez-Fournon}, {Geier},
  {Guti{\'e}rrez}, {Delgado-Gonz{\'a}lez}, {Angel}, {K{\"o}nyves-T{\'o}th},
  {Marques-Chaves}, {Omand}, \& {Shirley}}]{2023TNSCR3050....1P}
{Poidevin}, F., {P{\'e}rez-Fournon}, I., {Geier}, S., {et~al.} 2023, Transient
  Name Server Classification Report, 2023-3050, 1

\bibitem[{{Ramsden} {et~al.}(2022){Ramsden}, {Lanning}, {Nicholl}, \&
  {McGee}}]{2022MNRAS.515.1146R}
{Ramsden}, P., {Lanning}, D., {Nicholl}, M., \& {McGee}, S.~L. 2022, \mnras,
  515, 1146, \dodoi{10.1093/mnras/stac1810}

\bibitem[{{Rees}(1988)}]{1988Natur.333..523R}
{Rees}, M.~J. 1988, \nat, 333, 523, \dodoi{10.1038/333523a0}

\bibitem[{{Roming} {et~al.}(2005){Roming}, {Kennedy}, {Mason}, {Nousek}, {Ahr},
  {Bingham}, {Broos}, {Carter}, {Hancock}, {Huckle}, {Hunsberger}, {Kawakami},
  {Killough}, {Koch}, {McLelland}, {Smith}, {Smith}, {Soto}, {Boyd},
  {Breeveld}, {Holland}, {Ivanushkina}, {Pryzby}, {Still}, \&
  {Stock}}]{2005SSRv..120...95R}
{Roming}, P. W.~A., {Kennedy}, T.~E., {Mason}, K.~O., {et~al.} 2005, \ssr, 120,
  95, \dodoi{10.1007/s11214-005-5095-4}

\bibitem[{{Schlafly} \& {Finkbeiner}(2011)}]{2011ApJ...737..103S}
{Schlafly}, E.~F., \& {Finkbeiner}, D.~P. 2011, \apj, 737, 103,
  \dodoi{10.1088/0004-637X/737/2/103}

\bibitem[{{Science Software Branch at STScI}(2012)}]{2012ascl.soft07011S}
{Science Software Branch at STScI}. 2012, {PyRAF: Python alternative for IRAF},
  Astrophysics Source Code Library, record ascl:1207.011

\bibitem[{{Shingles} {et~al.}(2021){Shingles}, {Smith}, {Young}, {Smartt},
  {Tonry}, {Denneau}, {Heinze}, {Weiland}, {Flewelling}, {Stalder},
  {Clocchiatti}, {F{\"o}rster}, {Pignata}, {Rest}, {Anderson}, {Stubbs}, \&
  {Erasmus}}]{2021TNSAN...7....1S}
{Shingles}, L., {Smith}, K.~W., {Young}, D.~R., {et~al.} 2021, Transient Name
  Server AstroNote, 7, 1

\bibitem[{{Short} {et~al.}(2020){Short}, {Nicholl}, {Lawrence}, {Gomez},
  {Arcavi}, {Wevers}, {Leloudas}, {Schulze}, {Anderson}, {Berger}, {Blanchard},
  {Burke}, {Castro Segura}, {Charalampopoulos}, {Chornock}, {Galbany},
  {Gromadzki}, {Herzog}, {Hiramatsu}, {Horne}, {Hosseinzadeh}, {Howell},
  {Ihanec}, {Inserra}, {Kankare}, {Maguire}, {McCully}, {M{\"u}ller Bravo},
  {Onori}, {Sollerman}, \& {Young}}]{2020MNRAS.498.4119S}
{Short}, P., {Nicholl}, M., {Lawrence}, A., {et~al.} 2020, \mnras, 498, 4119,
  \dodoi{10.1093/mnras/staa2065}

\bibitem[{{Somalwar} {et~al.}(2023){Somalwar}, {Ravi}, {Yao}, {Guolo},
  {Graham}, {Hammerstein}, {Lu}, {Nicholl}, {Sharma}, {Stein}, {van Velzen},
  {Bellm}, {Coughlin}, {Groom}, {Masci}, \& {Riddle}}]{2023arXiv231003782S}
{Somalwar}, J.~J., {Ravi}, V., {Yao}, Y., {et~al.} 2023, arXiv e-prints,
  arXiv:2310.03782, \dodoi{10.48550/arXiv.2310.03782}

\bibitem[{{Stetson}(2000)}]{2000PASP..112..925S}
{Stetson}, P.~B. 2000, \pasp, 112, 925, \dodoi{10.1086/316595}

\bibitem[{{Subrayan} {et~al.}(2023){Subrayan}, {Milisavljevic}, {Chornock},
  {Margutti}, {Alexander}, {Ramakrishnan}, {Duffell}, {Dickinson}, {Lee},
  {Giannios}, {Lentner}, {Linvill}, {Garretson}, {Graham}, {Stern},
  {Brethauer}, {Duong}, {Jacobson-Gal{\'a}n}, {LeBaron}, {Matthews}, {Sears},
  \& {Venkatraman}}]{2023ApJ...948L..19S}
{Subrayan}, B.~M., {Milisavljevic}, D., {Chornock}, R., {et~al.} 2023, \apjl,
  948, L19, \dodoi{10.3847/2041-8213/accf1a}

\bibitem[{{Tonry} {et~al.}(2018){Tonry}, {Denneau}, {Heinze}, {Stalder},
  {Smith}, {Smartt}, {Stubbs}, {Weiland}, \& {Rest}}]{2018PASP..130f4505T}
{Tonry}, J.~L., {Denneau}, L., {Heinze}, A.~N., {et~al.} 2018, \pasp, 130,
  064505, \dodoi{10.1088/1538-3873/aabadf}

\bibitem[{{Valenti} {et~al.}(2016){Valenti}, {Howell}, {Stritzinger}, {Graham},
  {Hosseinzadeh}, {Arcavi}, {Bildsten}, {Jerkstrand}, {McCully}, {Pastorello},
  {Piro}, {Sand}, {Smartt}, {Terreran}, {Baltay}, {Benetti}, {Brown},
  {Filippenko}, {Fraser}, {Rabinowitz}, {Sullivan}, \&
  {Yuan}}]{2016MNRAS.459.3939V}
{Valenti}, S., {Howell}, D.~A., {Stritzinger}, M.~D., {et~al.} 2016, \mnras,
  459, 3939, \dodoi{10.1093/mnras/stw870}

\bibitem[{{van Velzen}(2018)}]{2018ApJ...852...72V}
{van Velzen}, S. 2018, \apj, 852, 72, \dodoi{10.3847/1538-4357/aa998e}

\bibitem[{{van Velzen} \& {Farrar}(2014)}]{2014ApJ...792...53V}
{van Velzen}, S., \& {Farrar}, G.~R. 2014, \apj, 792, 53,
  \dodoi{10.1088/0004-637X/792/1/53}

\bibitem[{{van Velzen} {et~al.}(2021{\natexlab{a}}){van Velzen}, {Pasham},
  {Komossa}, {Yan}, \& {Kara}}]{2021SSRv..217...63V}
{van Velzen}, S., {Pasham}, D.~R., {Komossa}, S., {Yan}, L., \& {Kara}, E.~A.
  2021{\natexlab{a}}, \ssr, 217, 63, \dodoi{10.1007/s11214-021-00835-6}

\bibitem[{{van Velzen} {et~al.}(2021{\natexlab{b}}){van Velzen}, {Gezari},
  {Hammerstein}, {Roth}, {Frederick}, {Ward}, {Hung}, {Cenko}, {Stein},
  {Perley}, {Taggart}, {Foley}, {Sollerman}, {Blagorodnova}, {Andreoni},
  {Bellm}, {Brinnel}, {De}, {Dekany}, {Feeney}, {Fremling}, {Giomi}, {Golkhou},
  {Graham}, {Ho}, {Kasliwal}, {Kilpatrick}, {Kulkarni}, {Kupfer}, {Laher},
  {Mahabal}, {Masci}, {Miller}, {Nordin}, {Riddle}, {Rusholme}, {van Santen},
  {Sharma}, {Shupe}, \& {Soumagnac}}]{2021ApJ...908....4V}
{van Velzen}, S., {Gezari}, S., {Hammerstein}, E., {et~al.} 2021{\natexlab{b}},
  \apj, 908, 4, \dodoi{10.3847/1538-4357/abc258}

\bibitem[{Virtanen {et~al.}(2020)Virtanen, Gommers, Oliphant, Haberland, Reddy,
  Cournapeau, Burovski, Peterson, Weckesser, Bright, {van der Walt}, Brett,
  Wilson, Millman, Mayorov, Nelson, Jones, Kern, Larson, Carey, Polat, Feng,
  Moore, {VanderPlas}, Laxalde, Perktold, Cimrman, Henriksen, Quintero, Harris,
  Archibald, Ribeiro, Pedregosa, {van Mulbregt}, \& {SciPy 1.0
  Contributors}}]{2020SciPy-NMeth}
Virtanen, P., Gommers, R., Oliphant, T.~E., {et~al.} 2020, Nature Methods, 17,
  261, \dodoi{10.1038/s41592-019-0686-2}

\bibitem[{{Wevers} {et~al.}(2022){Wevers}, {Nicholl}, {Guolo},
  {Charalampopoulos}, {Gromadzki}, {Reynolds}, {Kankare}, {Leloudas},
  {Anderson}, {Arcavi}, {Cannizzaro}, {Chen}, {Ihanec}, {Inserra},
  {Guti{\'e}rrez}, {Jonker}, {Lawrence}, {Magee}, {M{\"u}ller-Bravo}, {Onori},
  {Ridley}, {Schulze}, {Short}, {Hiramatsu}, {Newsome}, {Terwel}, {Yang}, \&
  {Young}}]{2022A&A...666A...6W}
{Wevers}, T., {Nicholl}, M., {Guolo}, M., {et~al.} 2022, \aap, 666, A6,
  \dodoi{10.1051/0004-6361/202142616}

\bibitem[{{Wiseman} {et~al.}(2024){Wiseman}, {Williams}, {Arcavi}, {Galbany},
  {Graham}, {H{\"o}nig}, {Newsome}, {Subrayan}, {Sullivan}, {Wang}, {Ili{\'c}},
  {Nicholl}, {Oates}, {Petrushevska}, \& {Smith}}]{2024arXiv240611552W}
{Wiseman}, P., {Williams}, R.~D., {Arcavi}, I., {et~al.} 2024, arXiv e-prints,
  arXiv:2406.11552, \dodoi{10.48550/arXiv.2406.11552}

\bibitem[{Wong {et~al.}(2022)Wong, Pfister, \& Dai}]{Wong_2022}
Wong, T. H.~T., Pfister, H., \& Dai, L. 2022, The Astrophysical Journal
  Letters, 927, L19, \dodoi{10.3847/2041-8213/ac5823}

\bibitem[{{Wyrzykowski} {et~al.}(2017){Wyrzykowski}, {Zieli{\'n}ski},
  {Kostrzewa-Rutkowska}, {Hamanowicz}, {Jonker}, {Arcavi}, {Guillochon},
  {Brown}, {Koz{\l}owski}, {Udalski}, {Szyma{\'n}ski}, {Soszy{\'n}ski},
  {Poleski}, {Pietrukowicz}, {Skowron}, {Mr{\'o}z}, {Ulaczyk}, {Pawlak},
  {Rybicki}, {Greiner}, {Kr{\"u}hler}, {Bolmer}, {Smartt}, {Maguire}, \&
  {Smith}}]{2017MNRAS.465L.114W}
{Wyrzykowski}, {\L}., {Zieli{\'n}ski}, M., {Kostrzewa-Rutkowska}, Z., {et~al.}
  2017, \mnras, 465, L114, \dodoi{10.1093/mnrasl/slw213}

\bibitem[{{Yao}(2023)}]{2023TNSCR2004....1Y}
{Yao}, Y. 2023, Transient Name Server Classification Report, 2023-2004, 1

\bibitem[{{Yao} {et~al.}(2021{\natexlab{a}}){Yao}, {Chu}, {Das}, {Kulkarni},
  {Somalwar}, {Gezari}, {Velzen}, \& {Hammerstein}}]{2021TNSCR3611....1Y}
{Yao}, Y., {Chu}, M., {Das}, K.~K., {et~al.} 2021{\natexlab{a}}, Transient Name
  Server Classification Report, 2021-3611, 1

\bibitem[{{Yao} {et~al.}(2021{\natexlab{b}}){Yao}, {Velzen}, {Perley},
  {Gezari}, {Hammerstein}, {Somalwar}, {Sharma}, \&
  {Kulkarni}}]{2021TNSCR1632....1Y}
{Yao}, Y., {Velzen}, S.~V., {Perley}, D., {et~al.} 2021{\natexlab{b}},
  Transient Name Server Classification Report, 2021-1632, 1

\bibitem[{{Yao} {et~al.}(2022){Yao}, {Lu}, {Guolo}, {Pasham}, {Gezari},
  {Gilfanov}, {Gendreau}, {Harrison}, {Cenko}, {Kulkarni}, {Miller}, {Walton},
  {Garc{\'\i}a}, {van Velzen}, {Alexander}, {Miller-Jones}, {Nicholl},
  {Hammerstein}, {Medvedev}, {Stern}, {Ravi}, {Sunyaev}, {Bloom}, {Graham},
  {Kool}, {Mahabal}, {Masci}, {Purdum}, {Rusholme}, {Sharma}, {Smith}, \&
  {Sollerman}}]{2022ApJ...937....8Y}
{Yao}, Y., {Lu}, W., {Guolo}, M., {et~al.} 2022, \apj, 937, 8,
  \dodoi{10.3847/1538-4357/ac898a}

\bibitem[{{Yao} {et~al.}(2023){Yao}, {Ravi}, {Gezari}, {van Velzen}, {Lu},
  {Schulze}, {Somalwar}, {Kulkarni}, {Hammerstein}, {Nicholl}, {Graham},
  {Perley}, {Cenko}, {Stein}, {Ricarte}, {Chadayammuri}, {Quataert}, {Bellm},
  {Bloom}, {Dekany}, {Drake}, {Groom}, {Mahabal}, {Prince}, {Riddle},
  {Rusholme}, {Sharma}, {Sollerman}, \& {Yan}}]{2023ApJ...955L...6Y}
{Yao}, Y., {Ravi}, V., {Gezari}, S., {et~al.} 2023, \apjl, 955, L6,
  \dodoi{10.3847/2041-8213/acf216}

\bibitem[{{Yaron} {et~al.}(2021){Yaron}, {Sass}, {Gal-Yam}, \&
  {Knezevic}}]{2021TNSAN.142....1Y}
{Yaron}, O., {Sass}, A., {Gal-Yam}, A., \& {Knezevic}, N. 2021, Transient Name
  Server AstroNote, 142, 1

\bibitem[{{Zackay} {et~al.}(2016){Zackay}, {Ofek}, \&
  {Gal-Yam}}]{2016ApJ...830...27Z}
{Zackay}, B., {Ofek}, E.~O., \& {Gal-Yam}, A. 2016, \apj, 830, 27,
  \dodoi{10.3847/0004-637X/830/1/27}

\bibitem[{{Zauderer} {et~al.}(2013){Zauderer}, {Berger}, {Margutti}, {Pooley},
  {Sari}, {Soderberg}, {Brunthaler}, \& {Bietenholz}}]{p2}
{Zauderer}, B.~A., {Berger}, E., {Margutti}, R., {et~al.} 2013, \apj, 767, 152,
  \dodoi{10.1088/0004-637X/767/2/152}

\bibitem[{{Zhou} {et~al.}(2021){Zhou}, {Liu}, {Komossa}, {Cao}, {Ho}, {Chen},
  \& {Li}}]{2021ApJ...907...77Z}
{Zhou}, Z.~Q., {Liu}, F.~K., {Komossa}, S., {et~al.} 2021, \apj, 907, 77,
  \dodoi{10.3847/1538-4357/abcccb}

\end{thebibliography}
\bibliographystyle{aasjournal}

\end{document}